%% file: main.tex
\documentclass[acmsmall]{acmart}

\PassOptionsToPackage{hyphens}{url}
\usepackage{url}
\usepackage{soul}
\usepackage{xcolor}
\usepackage{comment}
\usepackage{enumitem}
\usepackage{subcaption}
\usepackage{siunitx}

\AtBeginDocument{%
  \providecommand\BibTeX{{%
    \normalfont B\kern-0.5em{\scshape i\kern-0.25em b}\kern-0.8em\TeX}}}

\newcommand{\subr}[1]{{\small\texttt{r/#1}}}
\newcommand{\dsname}[1]{{\small \texttt{#1}}}
\newcommand{\ampsep}{\,{\scriptsize\&}\,}

\newcommand{\revision}[1]{{\textcolor{black}{#1}}}

\setcopyright{acmcopyright}
\acmJournal{PACMHCI}
\acmYear{2022} \acmVolume{6} \acmNumber{CSCW2} \acmArticle{526} \acmMonth{11} \acmPrice{15.00}\acmDOI{10.1145/3555639}

\begin{document}

\title[Assessing Community Effects of Interventions on r/The\_Donald]{Make Reddit Great Again: Assessing Community Effects of Moderation Interventions on r/The\_Donald}


\author{Amaury Trujillo}
\email{amaury.trujillo@iit.cnr.it}
\orcid{0000-0001-6227-0944}
\author{Stefano Cresci}
\email{stefano.cresci@iit.cnr.it}
\orcid{0000-0003-0170-2445}
\affiliation{%
    \institution{Institute for Informatics and Telematics, National Research Council (IIT-CNR)}
    \streetaddress{via G. Moruzzi 1}
    \city{Pisa}
    \country{Italy}
    \postcode{56124}
}

\renewcommand{\shortauthors}{Trujillo and Cresci}


\begin{abstract}
The subreddit \subr{The\_Donald} was repeatedly denounced as a toxic and misbehaving online community, reasons for which it faced a sequence of moderation interventions by Reddit administrators. It was quarantined in June 2019, restricted in February 2020, and finally banned in June 2020, but despite precursory work on the matter, the effects of this sequence of interventions are still unclear. In this work, we follow a multidimensional causal inference approach, with data containing more than 15M posts made in a time frame of 2 years, to examine the effects of such interventions inside and outside of the subreddit. We find that the interventions greatly reduced the activity of problematic users. 
However, the interventions also caused an increase in toxicity and led users to share more polarized and less factual news. 
In addition, the restriction had stronger effects than the quarantine, and core users of \subr{The\_Donald} suffered stronger effects than the rest of users. Overall, our results provide evidence that the interventions had mixed effects and paint a nuanced picture of the consequences of community-level moderation strategies. We conclude by reflecting on the challenges of policing online platforms and on the implications for the design and deployment of moderation interventions.

\end{abstract}



\maketitle

\input{introduction}
\input{related-works}
\input{datasets}
\input{methods}
\input{results}
\input{discussion}
\input{conclusions}

\bibliographystyle{ACM-Reference-Format}
\bibliography{references}

\received{January 2022}
\received[revised]{April 2022}
\received[accepted]{August 2022}

\end{document}

%% file: introduction.tex
\section{Introduction}
\label{sec:introduction}
The social media and news aggregator platform Reddit is among the most popular Internet websites, ranking as the sixth most visited website and the third most visited social media in the United States, as of April 2022.\footnote{\url{https://www.alexa.com/topsites/countries/US}} The platform is organized in communities called subreddits, in which users submit and discuss content regarding the community's shared topics and interests. 
Subreddits cover nearly every aspect of life, including news, sports, science, technology, religion, and a broad spectrum of social and other activities. Overall, \emph{politics} is one of the most discussed topics on the platform, with several subreddits related to US politics consistently ranking among the most popular communities on the platform. 
The outreach of these communities is enormous, reaching millions of users on Reddit, other online platforms (e.g., Facebook, Twitter)~\cite{zannettou2017web}, and even the audience of traditional media~\cite{soliman2019characterization}. For these reasons, political subreddits such as \subr{politics}, \subr{The\_Donald}, \subr{conservative}, and others, have received much scholarly attention~\cite{soliman2019characterization,horta2021platform,rajadesingan2020quick}.

In addition to platform-wise rules and policies, each subreddit sets its own behavior guidelines. Furthermore, unlike other platforms, users on Reddit can create and moderate their own subreddits in collaboration with others. 
As such, each community presents unique characteristics and develops its own habits, participation culture, and moderation rules~\cite{soliman2019characterization}. Occasionally, some communities accept and even encourage aggressive and harmful behaviors. When such communities repeatedly violate Reddit's policies, platform administrators (i.e., Reddit personnel) intervene to moderate the subreddit. Due to the sensitive and politicized nature of such behaviors, some readers might find upsetting some phrases used herein for their characterization.

\begin{figure}[t]
    \centering
    \includegraphics[width=1\textwidth]{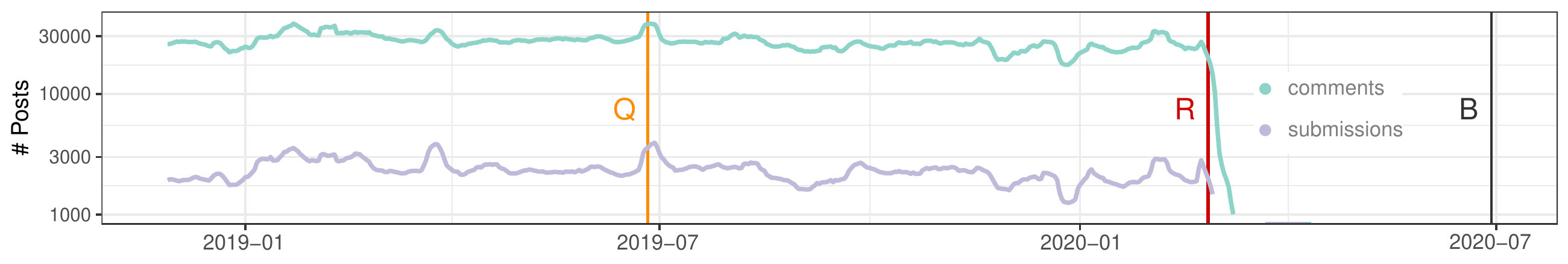}
    \caption{Seven-day running average of daily posts in \subr{The\_Donald} around quarantine~(Q), restriction~(R), and ban~(B). The number of submissions plummeted after the restriction, with comments following suit a few days later. Upon banning, the subreddit had already been completely inactive for several weeks.}
    \Description{A timeline chart shows the number of posts on a logarithmic scale for submissions and comments, both of which follow similar patterns up until the Restriction, with submissions halting almost immediately and comments following suit a few weeks afterward.}
    \label{fig:posts_time_series}
\end{figure}

\subsection{Moderating r/The\_Donald}
\label{sec:moderating-td}
The community of Donald Trump supporters of \subr{The\_Donald} was repeatedly denounced for toxicity, trolling, and harassment~\cite{flores2018mobilizing,massachs2020roots,mittos2020and,horta2021platform}.
For such misbehavior, Reddit administrators imposed a sequence of increasingly restrictive moderation interventions, as shown in Figure~\ref{fig:posts_time_series}:
\begin{enumerate}
    \item \textit{Quarantine (June 26, 2019)}: \subr{The\_Donald} was quarantined following repeated reports for inciting violence, including threatening US public figures. The subreddit was removed from the platform's search results and from the feed of non-subscribed users, albeit these could still access its content if they opted-in after receiving a warning message upon visit~\cite{chandrasekharan2020quarantined}. These measures intended to reduce the subreddit's visibility and to deter newcomers.
    \item \textit{Restriction (February 26, 2020)}: Administrators further restricted the subreddit and removed several of its moderators who were supporting content in violation of Reddit's policies, allowing new submissions to be posted only by approved users. Participation in the subreddit came to a complete halt within the following weeks, with the majority of users migrating to other subreddits or even to completely different platforms~\cite{horta2021platform}. The existing content remained accessible for reading and commenting after the restriction.
    \item \textit{Ban (June 29, 2020)}: \subr{The\_Donald} was banned, together with other two thousand subreddits, as part of Reddit's actions to enforce new policies. This ban permanently shut the subreddit down, by removing it from the platform and making it impossible to access its contents. 
\end{enumerate}

In addition to raising ethical and legal concerns as to whether online platforms should be allowed to limit the freedom of speech of their users, including that of major politicians~\cite{gillespie2018custodians,carlson2020you}, the practical consequences of these interventions are still unclear. Obviously, platforms apply moderation interventions to reduce the spread of toxic, hateful, fake, and otherwise problematic content~\cite{jhaver2021evaluating}. However, the extent to which such interventions are capable of mitigating the issues is still up to debate, and even more so given that certain interventions caused opposite effects to those planned ---i.e., they \textit{backfired}~\cite{chancellor2016thyghgapp,bail2018exposure,dias2020emphasizing,pennycook2020implied}. For these reasons, a growing body of studies has recently focused on evaluating social media moderation strategies~\cite{chandrasekharan2017you,saleem2018aftermath,chandrasekharan2020quarantined,horta2021platform,jhaver2021evaluating}.

While these early works provided interesting results, many questions remain unanswered. Firstly, interventions may produce effects spread across multiple dimensions of user behavior and ideology~\cite{lazer2018science}. For instance, they can affect user activity and participation to certain communities, political preferences, news consumption habits, and more. Early literature mainly investigated intervention effects with respect to user \textit{activity} and \textit{toxicity} (including hate speech)~\cite{chandrasekharan2017you,saleem2018aftermath,chandrasekharan2020quarantined,horta2021platform,jhaver2021evaluating}. However, many other important dimensions could be considered, such as the degrees of \textit{political polarization} and \textit{factual reporting}, which are known drivers of online misbehavior~\cite{bail2018exposure}. Currently, it is unknown whether past interventions produced any effect in these dimensions. Secondly, many different types of interventions are adopted by platform administrators, but the majority of existing studies only analyzed deplatforming interventions~\cite{chandrasekharan2017you,saleem2018aftermath,horta2021platform,jhaver2021evaluating} ---that is, interventions aimed at removing problematic users or communities from a platform, such as Reddit's restrictions and bans. The effects (or lack thereof) of other interventions, such as quarantines, are still unclear~\cite{chandrasekharan2020quarantined}. Thirdly, existing studies evaluated each intervention in isolation. However, certain interventions are enforced as part of a sequence of actions, as in the case of \subr{The\_Donald}. Finally, interventions that do not completely shut a community (e.g., quarantines, as opposed to bans) may produce effects both within and outside the moderated community, because most users browse and interact with multiple communities at the same time. Thus, when a community suffers a moderation intervention, the behavior of the members of that community might change not only within \textit{that} community, but also in \textit{other communities} in which they participate. Regarding Reddit's interventions on \subr{The\_Donald}, it would be interesting to go beyond existing results by evaluating the intervention effects outside \subr{The\_Donald}. Overall, additional analyses are needed to improve our understanding of the effects of recent moderation interventions.

\subsection{Research Questions}
The present study complements the existing literature in some of the aforementioned directions. 
In particular, Reddit's quarantines are almost completely unexplored, since the majority of existing studies focused on restrictions and bans~\cite{chandrasekharan2017you,saleem2018aftermath,horta2021platform,jhaver2021evaluating}. We extend and complement the only existing related study~\cite{chandrasekharan2020quarantined} by investigating the effects of Reddit's quarantine on \subr{The\_Donald} with respect to the toxicity of the affected users and by differentiating between different types of users. 
For this reason, we ask the following starting question:
\renewcommand\labelenumi{\bfseries{RQ\theenumi:}}
\begin{enumerate}[leftmargin=5\parindent]
    \item What were the effects of the quarantine, in terms of activity and toxicity, within \subr{The\_Donald}?
\end{enumerate}

Following from this question, and given that focusing only on a small set of metrics could be misleading, we expand current literature by measuring effects also with respect to the quality of the news articles consumed by members of a community, in order to gain deeper insights into the effects that moderation interventions actually have~\cite{bail2018exposure}. We thus seek answers to the following question:
\begin{enumerate}[leftmargin=5\parindent]
    \setcounter{enumi}{1}
    \item What were the effects of the quarantine, in terms of the quality of shared news articles, within \subr{The\_Donald}?
\end{enumerate}

Finally, previous work evaluated interventions in isolation, even when these were part of a sequence. Similarly, many existing works evaluated intervention effects only within the moderated community, but an intervention might have, for instance, positive local effects but negative global ones (e.g., on other communities, or on the platform as a whole)~\cite{copland2020reddit,horta2021platform}. 
Studying intervention effects on other communities allows a thorough understanding of its consequences. Hence, we ask one final question:
\begin{enumerate}[leftmargin=5\parindent]
    \setcounter{enumi}{2}
    \item What were the effects of the sequence of interventions applied to \subr{The\_Donald}, on the other communities to which \subr{The\_Donald} members participated?
\end{enumerate}




\subsection{Summary of Methods, Findings, and Implications}

\subsubsection{Materials and Methods}
Our observational study is based on Reddit data comprising more than 15M posts and spanning circa 2 years, collected from both \subr{The\_Donald} and other subreddits in which core users of the former participated.\footnote{The main data and code of the study are available at  \url{https://doi.org/10.5281/zenodo.6250576}} 
We first operationalized intervention effects for user activity, comment toxicity, the degrees of both political polarization and factual reporting of shared news articles, and group community proclivity. We then leveraged appropriate statistical tests to assess all measured effects, including causal inference methods such as interrupted time series (ITS) regression analysis and Bayesian structural time series (BSTS) modeling.


\subsubsection{Findings}
Regarding the effects of the quarantine on activity and toxicity within \subr{The\_Donald} (\textbf{RQ1}), we find a moderate decrease in user activity as well as a strong short-term decrease, but a strong long-term increase in toxicity. The quarantine also caused a mild decrease in the degree of factual reporting of the news articles shared within \subr{The\_Donald} (\textbf{RQ2}). No significant effect was found for the political polarization of shared news articles. Results for \textbf{RQ1} and \textbf{RQ2} also reveal that core users of \subr{The\_Donald} suffered stronger effects with respect to other users. Finally, the analysis of the effects that the quarantine and the restriction had outside of \subr{The\_Donald} (\textbf{RQ3}) reveal a decrease in user activity, a marked increase in toxicity, a decrease on the degree of factual reporting, and an increase in political polarization of shared news articles. Results for \textbf{RQ3} also reveal that the restriction caused stronger effects than the quarantine.


\subsubsection{Implications and Significance}
Our results highlight that the sequence of interventions enforced on \subr{The\_Donald} had mixed effects. Overall, our results and other recent findings~\cite{horta2021platform} partly question the positive judgments expressed in several previous works about the efficacy of Reddit's~\cite{chandrasekharan2017you,saleem2018aftermath,chandrasekharan2020quarantined} and Twitter's~\cite{jhaver2021evaluating} interventions. Furthermore, our nuanced results call for renewed efforts at evaluating the possible side effects of an intervention and the temporal variations of its effects. This research also contributes to building theories and methods~\cite{kiesler2012regulating,gillespie2018custodians} to inform platform administrators for the design and deployment of future moderation interventions.


%% file: related-works.tex
\section{Background and Related Work}
\label{sec:related-work}
We provide background information on \subr{The\_Donald} and the many issues emerged therein, and a critical discussion of the existing literature on the effects of past moderation interventions.

\subsection{The Rise and Fall of r/The\_Donald}
\label{sec:related-work-donald}
Between 2015 and 2020 \subr{The\_Donald} served as an online space for supporters of the businessman and former US president Donald Trump. It was created on June 27, 2015, following the announcement of Trump's presidential campaign, and soon after it gained widespread popularity among Trump enthusiasts, as well as among conservative and libertarian users~\cite{massachs2020roots}. At its peak, it counted almost 800K subscribers and it frequently ranked in the top-10  subreddits by activity.\footnote{\url{https://subredditstats.com/r/the_donald}} Due to their technical skills, organization and motivation~\cite{gaudette2020upvoting,shepherd2020gaming,jungherr2021populist}, members of \subr{The\_Donald} managed to exert a strong influence on the news discussed on other social media platforms, like Twitter~\cite{zannettou2017web,chandrasekharan2020quarantined}.

Initially, discussions on \subr{The\_Donald} mainly focused on Trump-related news, with the vast majority of the posted content supporting his candidancy and presidency. However, through time the subreddit slowly regressed to an alt-right bastion~\cite{soliman2019characterization,horta2021platform} and a hub for far-right extremism~\cite{chandrasekharan2020quarantined}. The offensive nature of the content posted on \subr{The\_Donald} and the aggressive behavior of its members frequently caused considerable controversy and turmoil. Through the years, Reddit users, journalists and scholars repeatedly denounced \subr{The\_Donald} for being toxic and violent~\cite{soliman2019characterization,horta2021platform}, racist, sexist and Islamophobic~\cite{gaudette2020upvoting,rieger2021assessing}, engaged in coordinated trolling and harassment~\cite{flores2018mobilizing}, in strategic manipulation~\cite{shepherd2020gaming}, and in the spread of conspiracy theories~\cite{massachs2020roots}. The archetype of \subr{The\_Donald}'s member was that of a white Christian male interested in conspiracy theories, firearms, and video games, and engaged in shocking and vitriolic humor~\cite{massachs2020roots}. 

Many of the aggressive and harmful behaviors described above were in clear violation of Reddit's policies. For this reason, between 2019 and 2020 the platform administrators applied three increasingly restrictive moderation interventions to \subr{The\_Donald}. The first of such interventions (i.e., the quarantine) applied concepts of design friction~\cite{cox2016design} in order to make it more difficult for casual users to enter, and to be exposed to the content of, the subreddit~\cite{chandrasekharan2020quarantined}. Apart from the added difficulties however, all of the content from \subr{The\_Donald} remained visible and all interactions with content and users remained possible. Conversely, the second intervention (i.e., the restriction) made it impossible for the vast majority of users to post new submissions to the subreddit and eventually resulted in a mass migration of users to a new platform~\cite{horta2021platform}. In practice, the restriction doomed \subr{The\_Donald}, even before the final ban that occurred four months later. Our present work is part of the ongoing stream of research that aims at assessing the effects of these moderation interventions.

\subsection{Evaluating the Effects of Moderation Interventions}
\label{sec:related-work-effects}
The many issues that currently affect online platforms ---including those described above, as well as the spread of mis- and disinformation~\cite{wardle2017information,dipietro2021new}; of propaganda and conspiracy theories~\cite{dasanmartino2020survey,paudel2021soros}; the rise of hateful, abusive, and coordinated inauthentic behavior~\cite{founta2018large,nizzoli2021coordinated}; and the misbehavior of social bots and trolls~\cite{cresci2020decade,zannettou2019let,mendoza2020bots}--- mandate the design and deployment of a multitude of moderation interventions. Given this picture, a fundamental question arises about the efficacy of such interventions for mitigating the existing issues. 

The recent body of works that evaluated Reddit's quarantines, restrictions, and bans represents the literature that is mostly related to our present study. \citeauthor{chandrasekharan2017you} as well as \citeauthor{saleem2018aftermath} evaluated the effects of the bans that targeted \subr{fatpeoplehate} and \subr{CoonTown} in 2015, two subreddits whose users were known for harassment~\cite{chandrasekharan2017you,saleem2018aftermath}. \citeauthor{chandrasekharan2017you} measured overall positive effects for the interventions. Specifically, they found that many former members of \subr{fatpeoplehate} and \subr{CoonTown} ceased using Reddit and that those who remained on the platform markedly decreased their hate speech usage. In addition, \citeauthor{saleem2018aftermath} found that the counter-actions taken by the former \subr{fatpeoplehate} members to circumvent the ban were short-lived and ineffective~\cite{saleem2018aftermath}. As it often happens with deplatforming, members who remained on Reddit after the bans ``migrated'' to other subreddits~\cite{newell2016user,horta2021platform}. Those subreddits, however, saw no significant changes in hate speech usage after the interventions~\cite{chandrasekharan2017you}. Nonetheless, the former members of \subr{CoonTown} more than doubled their posting activity when they migrated to \subr{The\_Donald}~\cite{chandrasekharan2020quarantined}. Reddit's quarantining of \subr{The\_Donald} and \subr{TheRedPill} were evaluated in~\cite{chandrasekharan2020quarantined}. Authors concluded that the quarantines made it more difficult to recruit new members to the moderated communities, but that the overall degree of misogyny and racism of their existing members remained unaffected. 

The previous studies shed light on (some of) the effects that Reddit's interventions had within Reddit itself. However, since such interventions caused many users to migrate to other platforms, those studies did not provide answers as to whether Reddit's bans made those problematic users «someone else's problem»~\cite{chandrasekharan2017you}. \citeauthor{horta2021platform} aimed to answer this question in the aftermath of Reddit's 2020 deplatforming of \subr{The\_Donald} and \subr{Incels}, whose former users migrated respectively to {\small\texttt{thedonald.win}} and {\small\texttt{incels.co}}~\cite{horta2021platform}. They found that both interventions significantly decreased activity on the new platforms, reducing the number of shared posts, active users, and newcomers. However, former users of \subr{The\_Donald} showed increases in toxicity and radicalization, supporting the hypothesis that the reduction in activity may have come at the expense of a more toxic and radical community~\cite{horta2021platform}. Besides Reddit,~\citeauthor{jhaver2021evaluating} evaluated Twitter's ban of three controversial influencers. They found that the deplatforming intervention significantly reduced the number of conversations about all three individuals and that their supporters decreased their overall activity and their degree of toxicity after the intervention~\cite{jhaver2021evaluating}. 

\subsubsection{Relation with Prior Work}
Results from the studies previously discussed surface a general consensus that deplatforming interventions lead to overall positive results, at least with respect to reducing activity. At the same time however, nearly all studies also reported a number of negative side effects. 
In light of these results, one of the most recent studies along this line of research underlined the importance of carrying out nuanced analyses in order to accurately assess the (sometimes mixed) effects of moderation interventions applied to complex online social systems~\cite{horta2021platform}.
Within this scientific context, our present work extends and complements existing studies by investigating a number of unexplored aspects. Firstly, we extend previous analyses on activity and toxicity by also investigating the quality of the news articles (i.e., in terms of their political polarization and the degree of factual reporting) shared and consumed ---important aspects that contribute to accurately understanding the consequences of moderation interventions, including their negative side effects~\cite{bail2018exposure}. Secondly, we compare the effects of subsequent interventions applied to \subr{The\_Donald}, thus contributing to analyze their relative effectiveness~\cite{chandrasekharan2020quarantined,horta2021platform}. 
Finally, we contribute to filling the scientific gap related to the analysis of Reddit's quarantines, whose effects were so far only analyzed by~\cite{chandrasekharan2020quarantined} within the moderated community. Here instead, we also evaluate the nuanced effects that quarantining \subr{The\_Donald} had on other communities on Reddit, similarly to what~\cite{horta2021platform} did when moderated communities migrated to other platforms.

%% file: datasets.tex
\input{tab-datasets}

\section{Data}
\label{sec:datasets}
Our study is based on observational Reddit data comprising more than 15M posts collected over the course of two years. In detail, our data is organized into three datasets that respectively contain: (i) all content shared within \subr{The\_Donald} in a time frame centered around the quarantine; (ii) all content shared within \subr{The\_Donald} by core users of the subreddit, in a time frame centered around the quarantine; and (iii) all content shared outside of \subr{The\_Donald} (i.e., in all other subreddits) by core users of \subr{The\_Donald}, in a large time frame that encompasses both quarantine and restriction. The mapping of our datasets to our research questions is shown in Table~\ref{tab:datasets}, together with other summary information. In detail, we use the first and second dataset to answer to \textbf{RQ1} and \textbf{RQ2} (quarantine effects within \subr{The\_Donald}). Instead, we use the third dataset to answer to \textbf{RQ3} (quarantine and restriction effects outside of \subr{The\_Donald}).

Posts on Reddit can be either submissions or comments. 
These represent two intrinsically different activities that, in general, follow different dynamics~\cite{weninger2014exploration}. Moreover, certain moderation interventions (e.g., the restriction suffered by \subr{The\_Donald}) are designed to have a strong impact on submissions, while leaving commenting activities unaffected. We thus conducted separate analyses for submissions and comments. 

\subsection{Data Source and Time Frame}
\label{sec:datasets-time-frame}
\begin{figure}[t]
    \centering
    \includegraphics[width=1\textwidth]{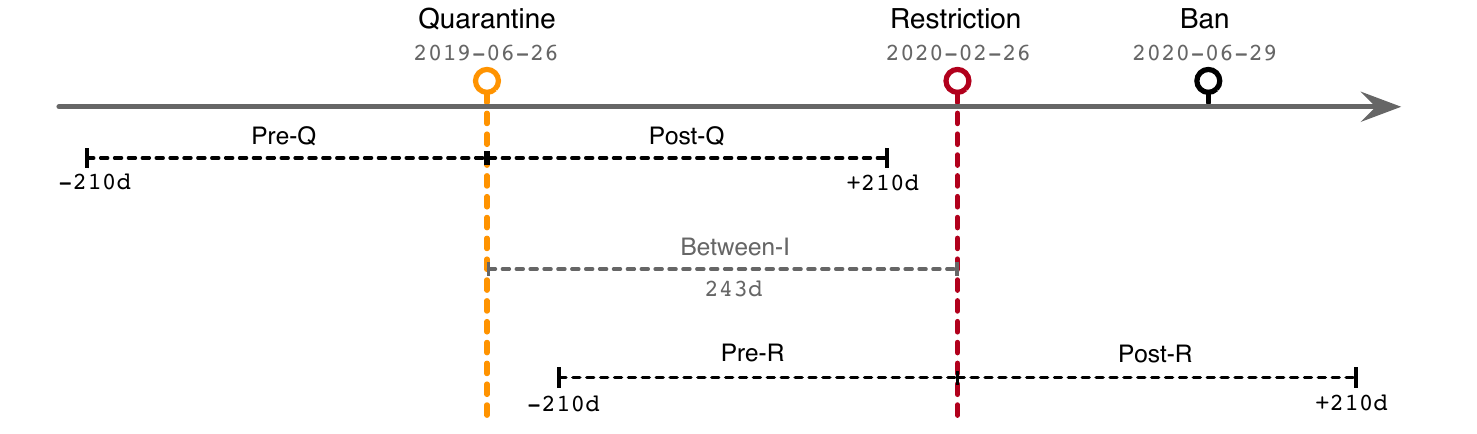}
    \caption{Timeline depicting the pre-post periods used for data collection and analysis, based on 2 time windows centered around the quarantine and the restriction, plus a convenience period between both interventions.}
    \Description{A timeline of interventions (quarantine, restriction, and ban) is shown, with two time windows centered around the first two interventions, in which each pre-post intervention period consists of 210 days. Between the quarantine and restriction there is a slightly larger period of 243 days that covers both the post-quarantine and pre-restriction periods.}
    \label{fig:interventions-timeline}
\end{figure}

For the collection of all Reddit data used herein, including the posts from \subr{The\_Donald} that are no longer available in Reddit because of the ban, we used the monthly archives from Pushshift, a service that provides historical Reddit data~\cite{baumgartner2020pushshift}. As already shown in Figure~\ref{fig:posts_time_series} in \S\ref{sec:moderating-td}, daily posting activity in \subr{The\_Donald} came to a halt shortly after the restriction, thus the subreddit was already inactive when the ban occurred. Consequently, and in line with previous work~\cite{horta2021platform}, the time frame considered for our data collection and analyses is centered around the quarantine and the restriction. 
In literature, different time frames were used: ±60 days around interventions~\cite{saleem2018aftermath}, ±120 days~\cite{horta2021platform}, ±180 days~\cite{chandrasekharan2020quarantined, jhaver2021evaluating}, and ±200 days~\cite{chandrasekharan2017you}. Given that circa 245 days passed between the quarantine and the restriction, we used a time frame spanning ±210 days (30 weeks). This choice allows us to divide a given time window around an intervention by groups of ten days~\cite{chandrasekharan2017you, chandrasekharan2020quarantined} or seven days, which eases the analysis of our time series. 
We thus collected data from November 28, 2018 to September 23, 2020 (inclusive), in which 1.06M submissions and 12.3M comments were posted to \subr{The\_Donald}. We then further divided the daily content into two pre-post intervention periods around quarantine~(Q) and restriction~(R), plus a convenience period between both ---all of which exclude the intervention days--- as illustrated in Figure~\ref{fig:interventions-timeline}. 


\subsection{Core Users}
\label{sec:datasets-core-users}
The selection of representative members of a community (i.e., core users) is an inherently subjective ---but sometimes inevitable~\cite{nizzoli2021coordinated}--- process. Before the quarantine \subr{The\_Donald} was a public space in which any registered user could post. As a result, there are many users who participated in the subreddit very sporadically, or even only once during the time frame of our study. Moreover, several users authored many posts in \subr{The\_Donald}, but only on a single day, or for a single submission or thread. Keeping in mind that interventions can have different effects depending on user characteristics and that the interventions enforced by Reddit were targeted at the core members of \subr{The\_Donald}, we operationalized our definition of \textit{core users}. 
We identify core users and we define them as those users who authored at least one post (i.e., either a submission or comment) a week, for the whole 30 weeks of the pre-quarantine period. 
In this way, we ensure that each core user had a non-negligible posting activity and a prolonged involvement with the community (circa seven months), before any intervention took place. Based on this definition, and after removing a few moderation bot accounts, we identified 2,239 core users of \subr{The\_Donald}, which represent only 1.12\% of the 198.5K distinct post authors in the subreddit for the study time frame. Despite their small number, core users posted 37.5\% of the total submissions and more than 53.9\% of the total comments.

\subsection{Datasets Construction and Preliminary Analysis}
\label{sec:preliminary-analysis}
After the selection of the appropriate study time frame (\S\ref{sec:datasets-time-frame}) and of the core users (\S\ref{sec:datasets-core-users}), we constructed our three datasets, summarized in Table~\ref{tab:datasets}. The first dataset (\dsname{TD}) is centered around the quarantine, and contains \subr{The\_Donald} posts made by all users (982K submissions and 11.4M comments). The second dataset (\dsname{CUw/iTD}) is a subset of the previous, with only posts made by core users within \subr{The\_Donald} (233.8K submissions and 3.29M comments). The third dataset (\dsname{CUw/oTD}) contains all posts that core users made outside of \subr{The\_Donald}, independently on the subreddit. This last dataset encompasses both the quarantine and the restriction. 

In order to better inform our choice of inference methods we also conducted an exploratory time series analysis on daily content and unique users for all of the datasets. This step allowed us to test for autocorrelation and seasonality before any intervention (i.e., Pre-Q period), which can cause certain causal inference methods to yield inaccurate results~\cite{brodersen2015inferring}. For autocorrelation, we applied the Durbin-Watson (DW) test of the \texttt{lmtest} R package. To identify seasonality, we used iterative STL (seasonal and trend decomposition using {\small LOESS}) of the \texttt{forecast} R package, with which we also derive seasonal strength on a scale from zero to one. 
All of the time series analyzed showed a significant positive autocorrelation, with values for the DW statistic ranging from 0.582 to 1.536. We also identified a weekly seasonality, with a strength that ranges from modest to strong (0.181--0.764). Both general and core users of \subr{The\_Donald} were most active around mid-week and least active around the weekend during the Pre-Q period, albeit core users showed higher weekly seasonal strength. This seasonality also indicates that having a subdivision of time periods in groups of seven days is more appropriate than the ten days used in other works~\cite{chandrasekharan2017you, chandrasekharan2020quarantined}.

%% file: tab-datasets.tex
\begin{table}[t!]
    \centering
    \caption{Overview of the 3 datasets used to answer research questions (RQ) regarding the interventions (I) quarantine (Q) and restriction (R) on \subr{The\_Donald} (TD) and its core users (CU).}
    \label{tab:datasets}
    \begin{tabular}{lccrr}
    \toprule
    Dataset & RQ & I & Submissions & Comments \\
    \hline
    \dsname{TD}         & 1\ampsep{}2   & Q             & 981,980   & 11,400,674 \\
    \dsname{CUw/iTD}    & 1\ampsep{}2   & Q             & 233,789   &  3,293,273 \\
    \dsname{CUw/oTD}    & 3             & Q\ampsep{}R   & 148,054   &  3,191,170 \\
    \bottomrule
    \end{tabular}
\end{table}

%% file: methods.tex
\section{Methods}
\label{sec:method}
\begin{figure}[t]
    \centering
    \includegraphics[width=0.65\textwidth]{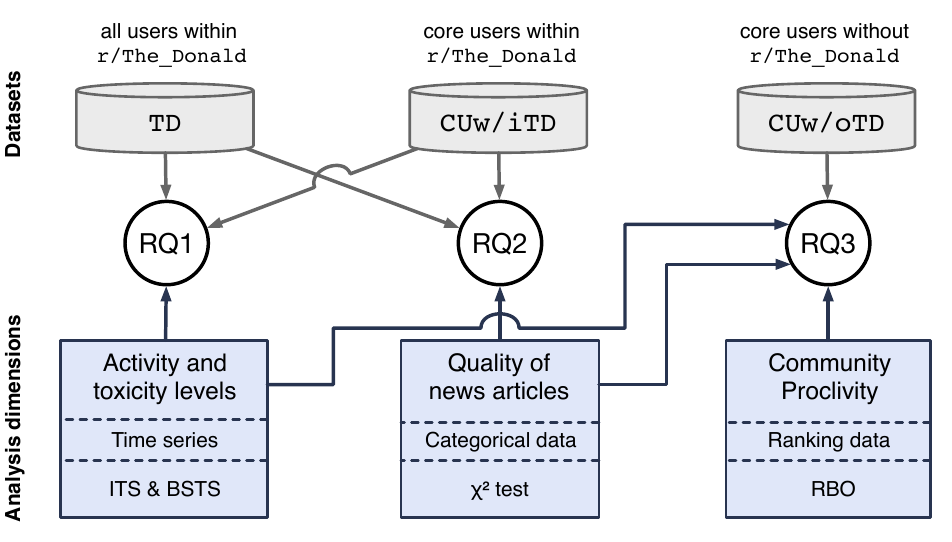}
    \caption{Conceptual approach used herein to investigate \subr{The\_Donald} interventions' community effects on activity and toxicity (RQ1), on the quality of its shared news articles (RQ2), and on other subreddits (RQ3).}
    \Description{Figure is composed of three main layers on a vertical basis. At the top we have three datasets: all users within The\_Donald, core users within the subreddit, and core users wihtout the subreddit. In the middle we have the three research questions of the study, each corresponding to a dataset in the previous layer. At the bottom we have the analysis dimensions used for the research questions: activity and toxicity levels (i.e., time series data) for RQ1 and RQ3; quality of news articles (i.e., categorical data) for RQ2; and group community proclivity (i.e., ranking data) for RQ3.}
    \label{fig:methodological_approach_diagram}
\end{figure}

Based on the nature of our datasets, we defined intervention effects, gathered additional data, and selected statistical descriptive and inference techniques apt to answer to our research questions. The conceptual approach we followed is illustrated in Figure~\ref{fig:methodological_approach_diagram} and detailed in the following paragraphs.

\subsection{Operationalizing Effects}
A single intervention can have multiple effects~\cite{lazer2018science}. Therefore, in order to accurately assess intervention effects it is necessary to investigate possible changes in many dimensions. 

\subsubsection{Activity} \label{sec:methods-activity}
We computed two separate content activity metrics based on Reddit's submissions and comments: (i) daily number of submissions and (ii) daily number of comments. Another commonly used activity metric is the number of daily active users (DAU). 
However, what constitutes an active user differs from platform to platform. We reprise the definition for core users given in \S\ref{sec:datasets-core-users} to define DAU as those users who performed at least a posting activity on a given day. Analogous to our metrics for posted content, we have (iii) submission DAU and (iv) comment DAU. These are thus the four metrics we use herein to measure intervention effects on activity.

\subsubsection{Toxicity}
Many previous studies evaluated the effectiveness of interventions for hate speech reduction~\cite{chandrasekharan2017you,saleem2018aftermath,chandrasekharan2020quarantined}. However, we lack a common definition of hate speech~\cite{davidson2017automated,rieger2021assessing}, its detection is often based on the presence of certain hateful words, with dictionary construction being challenging and context-dependent~\cite{davidson2017automated,fortuna2021well}, and the mere presence of hateful words has limited power due to the varied expressions of online incivility. For these reasons, others recently used toxicity as a more general concept to study moderation interventions~\cite{jhaver2021evaluating,horta2021platform} and as a discussion quality indicator~\cite{rajadesingan2020quick}. In this context, the Jigsaw unit at Google developed the Perspective API,\footnote{\url{https://www.perspectiveapi.com/}} a widely used public service that computes multiple toxicity-related scores for comments~\cite{rieder2021fabrics}.

Herein we rely on the \textit{severe toxicity} score of this service, similarly to several previous works~\cite{jhaver2021evaluating,horta2021platform}. Severe toxicity is defined as being «very hateful, aggressive, disrespectful [...] or otherwise very likely to make a user leave a discussion or give up on sharing their perspective», and is considered to be the most reliable and robust\footnote{The severe toxicity score in Perspective is specifically designed to avoid predicting benign usage of foul language as toxic.} indicator of toxicity among those provided by the Perspective API~\cite{rajadesingan2020quick,horta2021platform}. From  severe toxicity, defined in the $[0, 1]$ range, we compute two metrics of toxicity for Reddit comments: daily median and daily relative frequency. The first one is computed as the median of the severe toxicity scores of comments aggregated on a daily basis. For the second metric, we first convert scores into binary labels by considering as ``severely toxic'' all those comments whose score is above a certain threshold ($\geq$ 0.5), we then compute the fraction of severely toxic comments on a daily basis. We include both metrics in our analyses because they provide slightly different information and both have been recently used in studies on online toxicity~\cite{rajadesingan2020quick,jhaver2021evaluating,horta2021platform}. For the second metric we chose a score $\geq$ 0.5 because we consider it a good sensitive default for binary classification, it has been used in previous literature \cite{rajadesingan2020quick}, and results are qualitatively similar with higher scores used in other related work, such as $\geq$ 0.8~\cite{horta2021platform}.


\subsubsection{Political Polarization}
\label{sec:political_polarization}
The degree of polarization of a community is another important metric when investigating the ``health'' of online spaces, with several interventions being specifically designed to reduced online polarization~\cite{bail2018exposure}. Still, concerns have surfaced about the possibility that deplatforming interventions might actually increase the polarization of affected users~\cite{horta2021platform}. For this reason we also evaluate intervention effects in terms of the overall degree of political polarization of the affected users.
To measure political polarization, we adopt an approach based on the analysis of news articles shared by users. To this end, we leverage data from Media Bias/Fact Check (MBFC),\footnote{\url{https://mediabiasfactcheck.com/}} a widely-used platform that provides expert-curated fact checks and audit information about a large set of US media outlets~\cite{bovet2019influence}. Despite the limitations of using a single source for bias rating, the validity of using MBFC data to study political bias and factual reporting of Reddit submissions has been demonstrated via a large-scale study with hundreds of millions of posts across thousands of subreddits~\cite{weld2021political}. Therefore, we first use the general \textit{bias} in MBFC data, in which news websites are classified as \emph{pro-science} (factual and non-politically inclined), \emph{satire}, \emph{conspiracy/pseudoscience} (unverifiable information related to known conspiracies), \emph{fake} (deliberate attempts to publish hoaxes and/or disinformation for profit or influence), or \emph{politically biased}. For the latter, MBFC provides a Likert-like rating with five labels based on the US political spectrum, ranging from ``left'' to ``right'', which we then use to measure political polarization. In our analyses we compare the distribution of such labels for the articles shared by a community before and after a given intervention, treating general bias as categorical data and political bias as ordinal data.

\subsubsection{Factual Reporting}
Grounding comments on facts and sharing a common vision of reality with those with which we interact are key elements to achieve healthy online conversations~\cite{benkler2018network}. Because of this, scholars have long monitored the extent to which online communities share and consume news from sources known for adopting fact-based journalistic practices~\cite{bovet2019influence}. Herein, we evaluate the extent to which users affected by a moderation intervention change their behavior with respect to the degree of factuality of the news sources they share, by leveraging once again data from MBFC. Specifically, for each news outlet MBFC provides a \textit{factual reporting} Likert-like rating with six labels, ranging from ``very low'' to  ``very high'' factuality. Similarly to our analyses of political polarization, we use factual reporting to study intervention effects by comparing their distribution before and after a given intervention, treating the labels as ordinal data.

\subsubsection{Group Community Proclivity}
\label{sec:methods-proclivity}
\begin{figure}[t]
    \centering
    \includegraphics[width=1\textwidth]{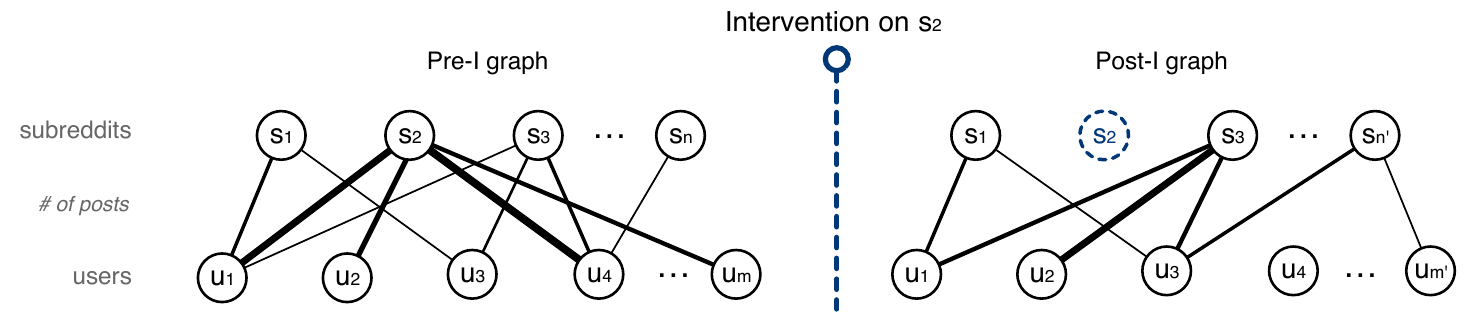}
    \caption{Representation of the user-subreddit relationship as a weighted undirected bipartite graph for the pre-post periods around an intervention on a given subreddit. As shown, an intervention on $s_2$ might induce users to alter their overall participation habits, thus altering the group community proclivity.}
    \Description{Figure is divided into two bipartite graphs. On the left, we have a pre-intervention graph composed of subreddits and users, with the number of posts being the weight of the edges that connect the two node kids. On the right we have the same graph after an intervention on the subreddit s2, which impedes creating edges between users and this particular subreddit, thus changing the overall connection structure of users and the rest of the subreddits.}
    \label{fig:user-subreddit-bipartite-graph}
\end{figure}

User participation in online communities can be represented not just by the number of posts or unique users (what we herein call activity), but also by the overall structural relationship between a group of users and the set of communities in a platform. That is, when a group of users ---as a whole--- is more inclined to participate in certain communities rather than others, the former are more important for that group. We refer to this concept as \emph{group community proclivity}. 
We are thus interested in comparing group community proclivity of core users of \subr{The\_Donald} within Reddit before and after interventions on the subreddit. Since we carry out separate analyses for submissions and comments, we have two bipartite graphs for each analyzed period. Our approach allows us to compare structural changes in the user-subreddit relationship before and after a given intervention, as illustrated in Figure~\ref{fig:user-subreddit-bipartite-graph}.

To obtain a metric of group community proclivity, we first represent the user-subreddit relationship for a given period as a weighted undirected bipartite graph $G = (U,S,E,w)$, where $U$ is the set of core users, $S$ is the set of subreddits, $E$ is the set of edges connecting nodes in $U$ and $S$, and $w$ are the weights of edges. An edge $e_{ij} \in E$ exists from $u_i \in U$ to $s_j \in S$ if $u_i$ has authored content on $s_j$ in the given period, with $w_{ij} \in \mathbb{N^+}$ being the number of posts authored by $u_i$ in $s_j$. We then adopt a ranking algorithm for bipartite graphs so as to obtain a ranking of subreddits according to their importance for the group of users. Among the available ranking algorithms, we adopted Co-HITS ---a version of the well-known HITS algorithm adapted to bipartite graphs~\cite{deng2009generalized}. In our analyses we used the Co-HITS implementation from the \texttt{birankr} R package. 


\subsection{Descriptive and Inference Methods}
\label{sec:inference-methods}
We now describe our choice of statistical methods for assessing intervention effects.

\subsubsection{RQ1: Effects on Activity and Toxicity within r/The\_Donald}
\label{sec:inference-methods-rq1}
We followed a quasi-experimental approach to measure the intervention effects on activity and toxicity within \subr{The\_Donald}. 
Based on our preliminary time series analysis (§\ref{sec:preliminary-analysis}), we leverage two causal inference methods to analyze possible intervention effects in a complementary manner: interrupted time series (ITS) regression analysis to describe the trend, onset and decay of intervention effects; and Bayesian structural time series (BSTS) modeling to compute effect size, confidence intervals, and significance.  

ITS regression analysis aims to establish the underlying trend of the variable of interest across a continuous sequence of observations before and after being ``interrupted'' by an intervention at a well-defined point in time. In its most basic form, ITS can be implemented through simple segmented linear regression models, which renders it easy to use, understand, and plot. For these reasons it was used in works on social media interventions, either as complementary to simpler methods such as difference-in-differences~\cite{chandrasekharan2017you}, or as the only causal inference method~\cite{chandrasekharan2020quarantined,jhaver2021evaluating}. However, specific regression families should be used in the case of non-linear trends and for particular underlying data distributions ~\cite{bernal2017interrupted}. Moreover, ITS has caveats in case of autocorrelation and seasonality, which demand adequate model adjustments so as to avoid misleading estimates of effect sizes, using for instance apt autoregressive integrated moving average (ARIMA) models~\cite{schaffer2021interrupted}, or heteroscedasticity and autocorrelation consistent (HAC) estimators~\cite[\S{15.4}]{hanck2019introduction}.
Herein we use a simple and interpretable ITS regression to visualize variations in the linear trends of the variables of interest around interventions, according to the following segmented linear model,
\begin{equation} \label{eq:segmented_lm}
Y_t = \beta_0 + \beta_1 T + \beta_2 D + \beta_3 P + \epsilon
\end{equation}
where $Y_t$ is the outcome variable of the time series; $T$ is a continuous variables that indicates the time in days from the start of the observational period, with $\beta_1$ indicating the trend before the intervention; $D$ is a dummy variable indicating the presence (1) or absence (0) of the intervention, with $\beta_2$ indicating the immediate change upon intervention; $P$ is a continuous variable that indicates the days passed since the intervention has occurred (from 0 to 210), with $\beta_3$ indicating the change in trend after intervention; finally, $\epsilon$ is the error term of the model. 

BSTS, on the other hand, is an approach based on Bayesian statistics that uses a structural time series model to capture the trend, seasonal, and related components of a time series, together with a dynamic regression component using Monte Carlo Markov Chain (MCMC) to create counterfactual data and confidence intervals~\cite{scott2014predicting}. 
BSTS improves on ITS in two main aspects: it provides a fully Bayesian estimate for the effect across time, which can be updated as new information is available; and it uses model averaging to construct a synthetic control to model the counterfactual. This means that the model's certainty will increase in case additional information is available, and that we can create a useful model even in the absence of an external control group. This is particularly valuable for our analyses due to the difficulties in finding an independent subreddit to act as a control. 
Indeed, Reddit users are mostly free to participate in as many subreddits at the same time as they wish, with most doing so, especially across related communities. This results in a significant user overlap in many communities, which thus are not independent from each other. Additionally, before the quarantine \subr{The\_Donald} was one of the most active subreddits, despite its focus on a single political figure. Finding a ``control'' subreddit with similar characteristics, let alone an independent one, would be problematic~\cite{chandrasekharan2020quarantined}. 
For these reasons, and despite the adoption of ITS in many recent studies that measured intervention effects~\cite{chandrasekharan2017you,chandrasekharan2020quarantined,jhaver2021evaluating}, BSTS represents a better alternative for estimating effects, given also the autocorrelation and weekly seasonality of our data. For our analyses we use the BSTS implementation of the \texttt{CausalImpact} R package. 


\subsubsection{RQ2: Effects on the Quality of Shared News Articles within r/The\_Donald}
\label{sec:methods-rq2}
Concerning the possible changes in the quality of shared news articles in \subr{The\_Donald}, we compare the political polarization and factual reporting scores of the news outlets linked in Reddit submissions, before and after a given intervention. Considered that only circa half of the collected submissions contain a link to an external website, of which only a subset point to news outlets in the MBFC database, we aggregate these links by pre- and post-intervention periods. 
To probe effects in political polarization, we further aggregate data on the left and right side of the political spectrum (excluding neutral news outlets, which are assigned the ``least biased'' label by MBFC). Then, we compare the differences in the distributions of political polarization scores before and after an intervention. Finally, we apply $\chi^2$ tests for significance, albeit with the awareness that the partially-matched data between interventions could give unreliable results in case of small differences. For estimating effects in factual reporting, we follow a similar approach where we first aggregate data by lower and upper ends on the factuality spectrum, and then we conduct $\chi^2$ tests on the proportions.

\subsubsection{RQ3: Effects of Multiple Interventions outside of r/The\_Donald}
\label{sec:methods-rq3}
To answer \textbf{RQ3}, we follow the methods to measure the possible effects on user activity and toxicity described for \textbf{RQ1}, as well as the methods for quality of shared news articles described for \textbf{RQ2}. In addition, we also evaluate the effects in group community proclivity (i.e., ranking variations computed with Co-HITS). 
In line with the approach used in RQ2, to assess group community proclivity we aggregate data on three periods: pre-quarantine, between-interventions, and post-restriction. Between-interventions is a convenience period that covers post-quarantine and pre-restriction, which we confirmed bears no significant difference with respect to the periods it substitutes, as there is less data compared to \emph{activity} and their spans mostly overlap, as sketched in Figure~\ref{fig:interventions-timeline}. To measure proclivity change between two periods, we use the rank-biased overlap (RBO) for indefinite lists~\cite{webber2010similarity}. RBO is a probabilistic model of similarity between two ranking lists based on average overlap, with a bias in the proportional overlap at each depth of weights, which can handle tied ranks and rankings of different lengths~\cite{webber2010similarity}. The latter is important in our case as other methods used for ranking similarity (e.g., Kendall's tau distance) only handle lists with the same items. However, with time some subreddits become inactive and new ones are created. In addition, we are particularly interested in changes about the top subreddits with the most proclivity, not on whole lists. We compute RBO scores with the corresponding function of the \texttt{gespeR} R package. 


%% file: results.tex
\section{Results}
\label{sec:results}

\subsection{RQ1: Effects on Activity and Toxicity within r/The\_Donald}
\label{sec:rq1-results}



\begin{figure}[t]
    \subcaptionbox{Content activity (all users).\label{fig:td_q_posts}}
        {\includegraphics[width=0.495\textwidth]{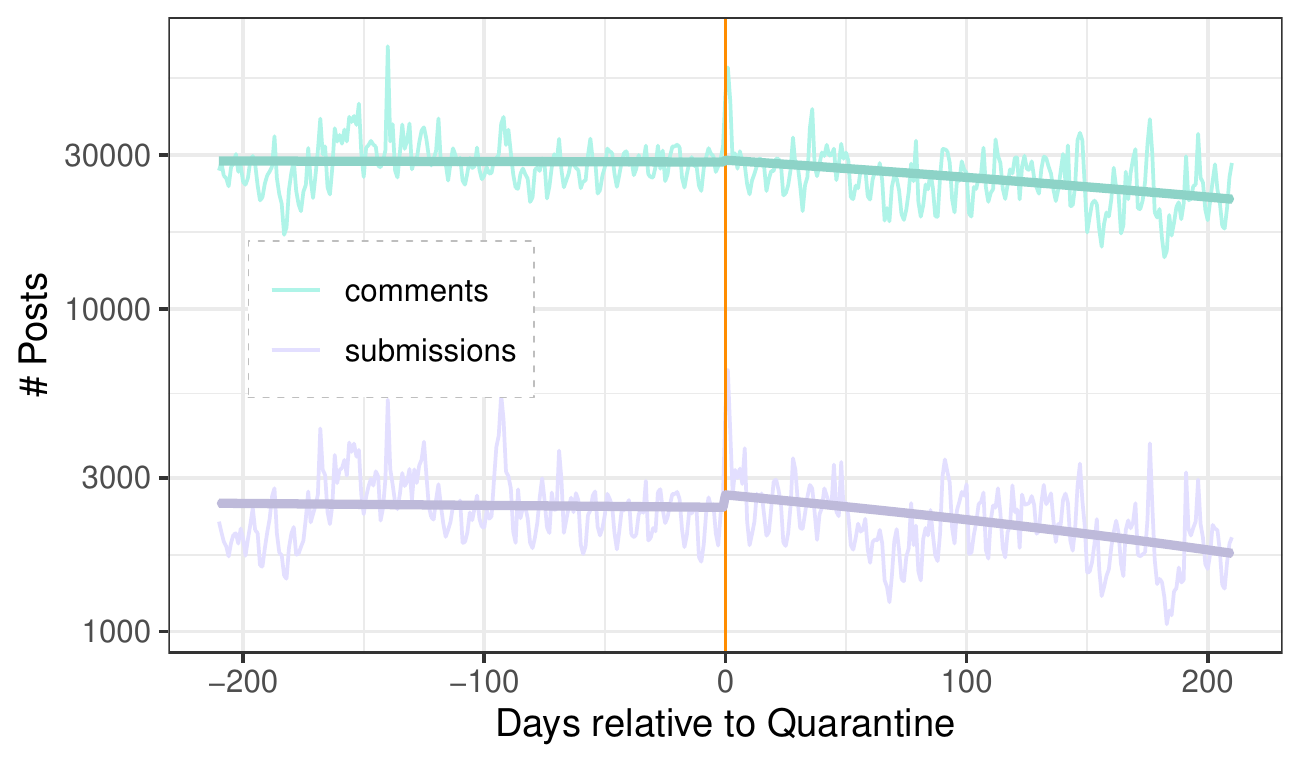}}
    \subcaptionbox{Content activity (core users).\label{fig:ctd_itd_q_posts}}
        {\includegraphics[width=0.495\textwidth]{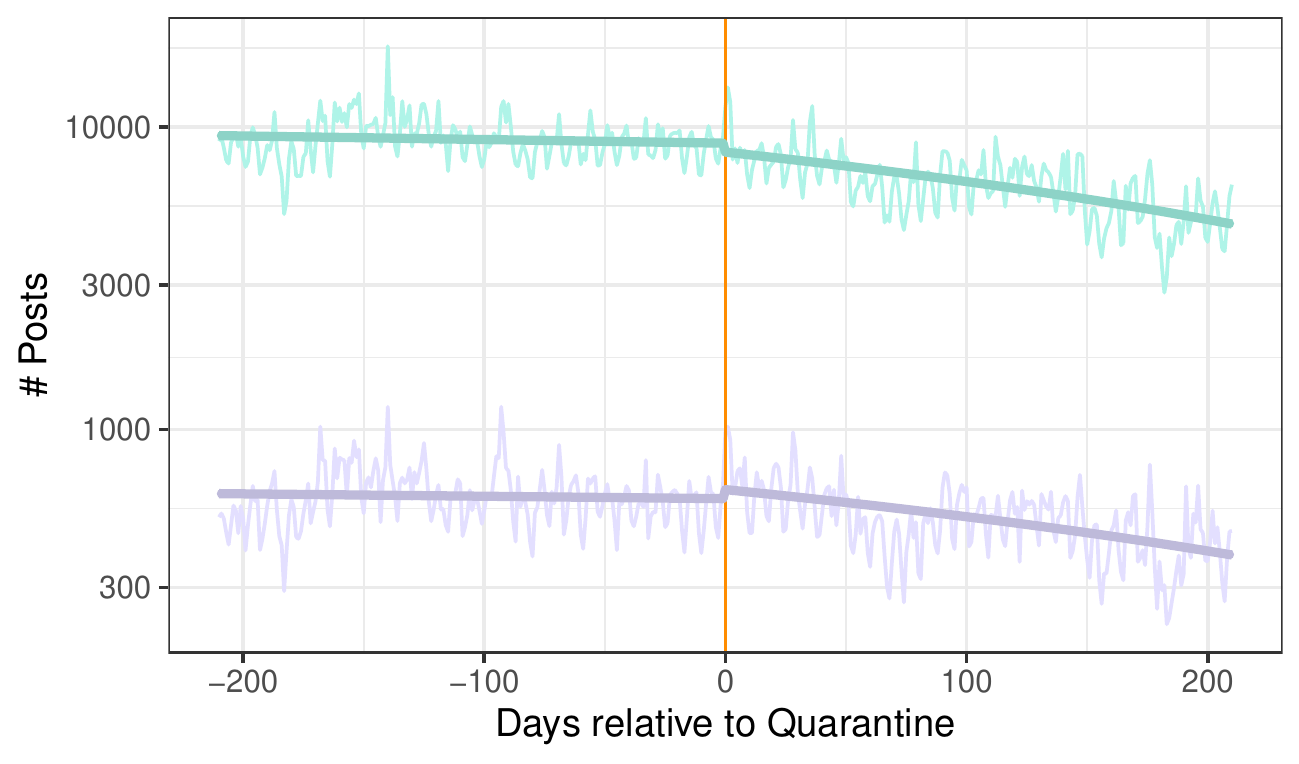}}
    \par\bigskip 
    \subcaptionbox{Daily active users (all users).\label{fig:td_q_dau}}
        {\includegraphics[width=0.495\textwidth]{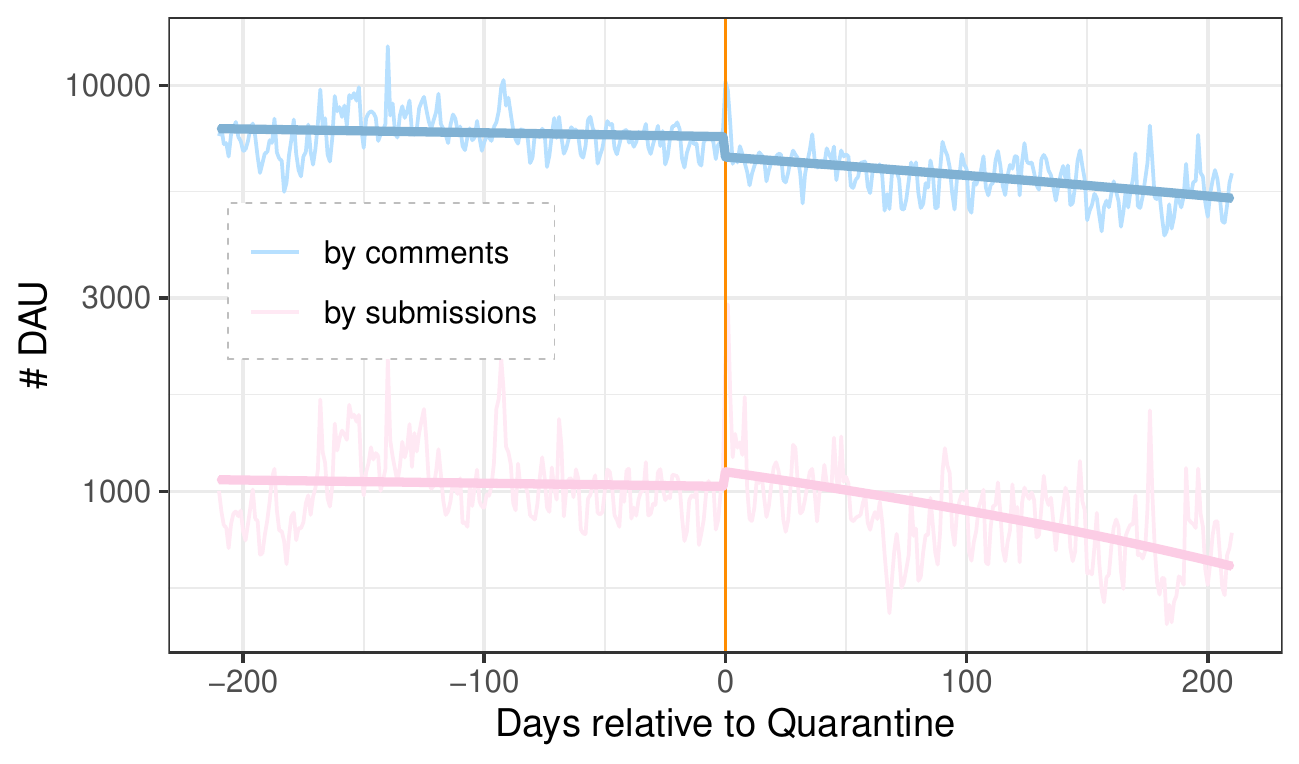}}
    \subcaptionbox{Daily active users (core users).\label{fig:ctd_itd_q_dau}}
        {\includegraphics[width=0.495\textwidth]{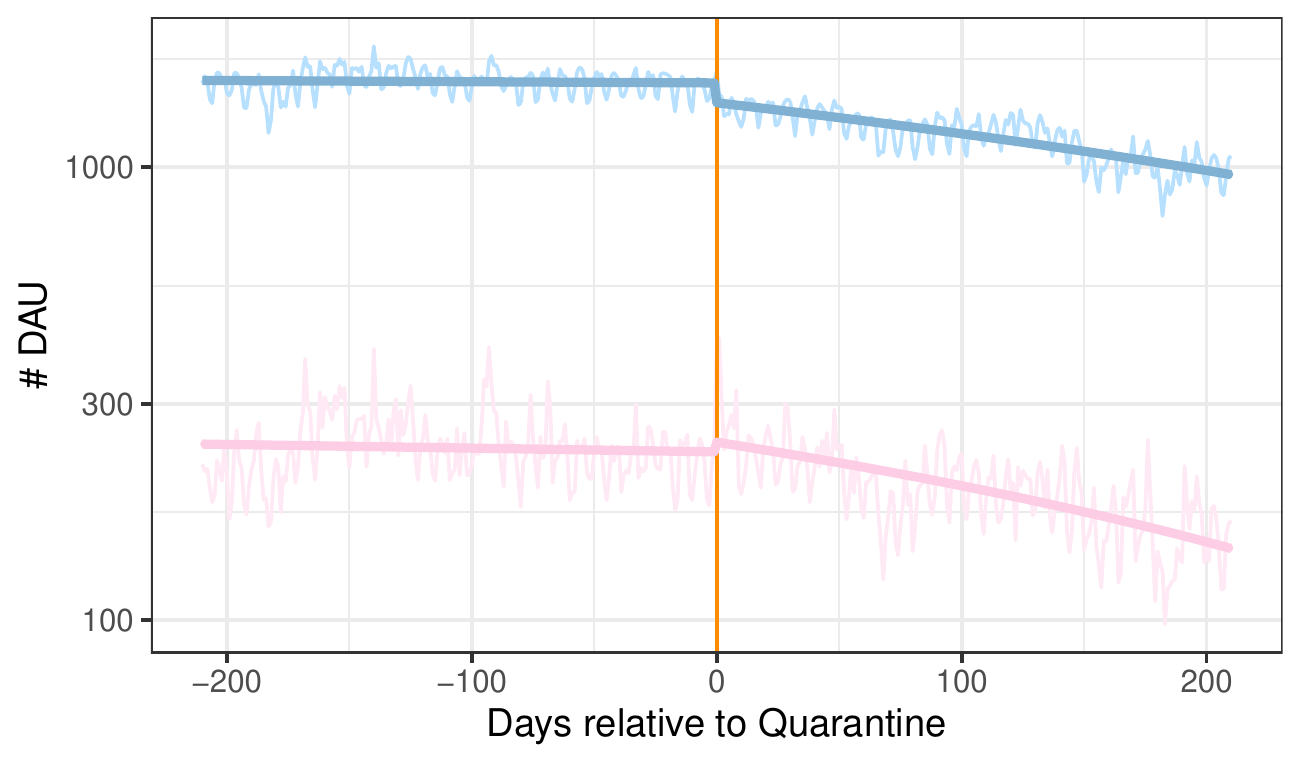}}
    \caption{ITS regressions of activity within \subr{The\_Donald} around quarantine, for all users and core users.}
    \Description{The content activity for all uses and core users suffers a modest decline after quarantine, with the number of daily active users by submissions being the most affected metric.}
    \label{fig:rq1_q_activity}
\end{figure}

\subsubsection{Activity}
\label{sec:rq1-activity}
The ITS analysis (Eq.~\ref{eq:segmented_lm} in \S\ref{sec:inference-methods-rq1}) for pre-quarantine shows that the linear trend ($\beta_1$) of activity in \subr{The\_Donald} was stable ---albeit slightly decreasing--- in terms of daily submissions, comments, and active users, as shown in Figure~\ref{fig:rq1_q_activity}. Upon quarantine, there were activity spikes in all metrics that lasted circa two days, but in all cases there is a noticeable subsequent effect ($\beta_3$) of decline in activity. However, the ITS immediate effect ($\beta_2$) was heterogeneous in direction among the metrics. For instance, upon quarantine there was an immediate increase on submission DAU but a decrease on comment DAU, as visible in Figures~\ref{fig:td_q_dau} and~\ref{fig:ctd_itd_q_dau}. At first glance, ITS seems to indicate that core users had a more marked decline in activity compared to all of the users. The BSTS analysis, reported in Table~\ref{tab:rq1_activity_bsts}, shows that all activity metrics had significant negative relative effects (i.e., average difference between observed and predicted post-intervention values). The most important effects concerned the number of daily active core users (-23\%) and the respective comments (-27\%), whilst the least important were on posts made by all users (-8.5\% for submissions and -10.7\% for comments). Hence, the quarantine indeed had a highly significant effect of reducing the activity of core users within \subr{The\_Donald}.

\begin{table}[t]
    \centering
    \caption{BSTS results of quarantine effects on activity within \subr{The\_Donald} for all users (\dsname{TD}) and core users (\dsname{CUw/iTD}). The lower the posterior tail-area probability (\textit{pp}), the higher the probability of a causal effect.}
    \label{tab:rq1_activity_bsts}
    \begin{tabular}{lrrcScc}
    \toprule
    & \multicolumn{3}{c}{Post-intervention avg. value} & \multicolumn{2}{c}{Relative effect (\%)} & \\
    \cmidrule(lr){2-4}\cmidrule(lr){5-6}
                                    & Actual  & Predicted & 95\% CI   & \multicolumn{1}{c}{\% Change} & 95\% CI & \textit{pp}\\ 
    \hline
    \dsname{TD} submissions     & 2201 & 2406 & [2215, 2609] & -8.5 & [-17, -0.57] & .019\\
    \dsname{TD} comments        & 25457 & 28501 & [27019, 29879]  & -10.7 & [-16, -5.5]  & .001\\
    \dsname{TD} submission {\small DAU}  & 886 & 1021 & [941, 1091] & -13.2 & [-20, -5.4] & .001\\
    \dsname{TD} comment {\small DAU}     & 5967 & 7492 & [7176, 7781]  & -20.4 & [-24, -16]  & .001\\
    \dsname{CUw/iTD} submissions& 509 & 589 & [545, 631] & -13.6 & [-21, -6.2] & .001\\
    \dsname{CUw/iTD} comments   & 6529 & 8940 & [8418, 9400]  & -27 & [-32, -21]  & .001\\
    \dsname{CUw/iTD} submission {\small DAU}  & 196 & 235 & [222, 248] & -16.7 & [-22, -11] & .001\\
    \dsname{CUw/iTD} comment {\small DAU}     & 1176 & 1529 & [1493, 1562]  & -23 & [-25, -21]  & .001\\
    \bottomrule
    \end{tabular}
\end{table}

\subsubsection{Toxicity}
\label{sec:rq1-toxicity}
\begin{figure}[t]
    \subcaptionbox{Median severe toxicity scores.\label{fig:ctd_itd_q_sevtox_median}}
        {\includegraphics[width=0.495\textwidth]{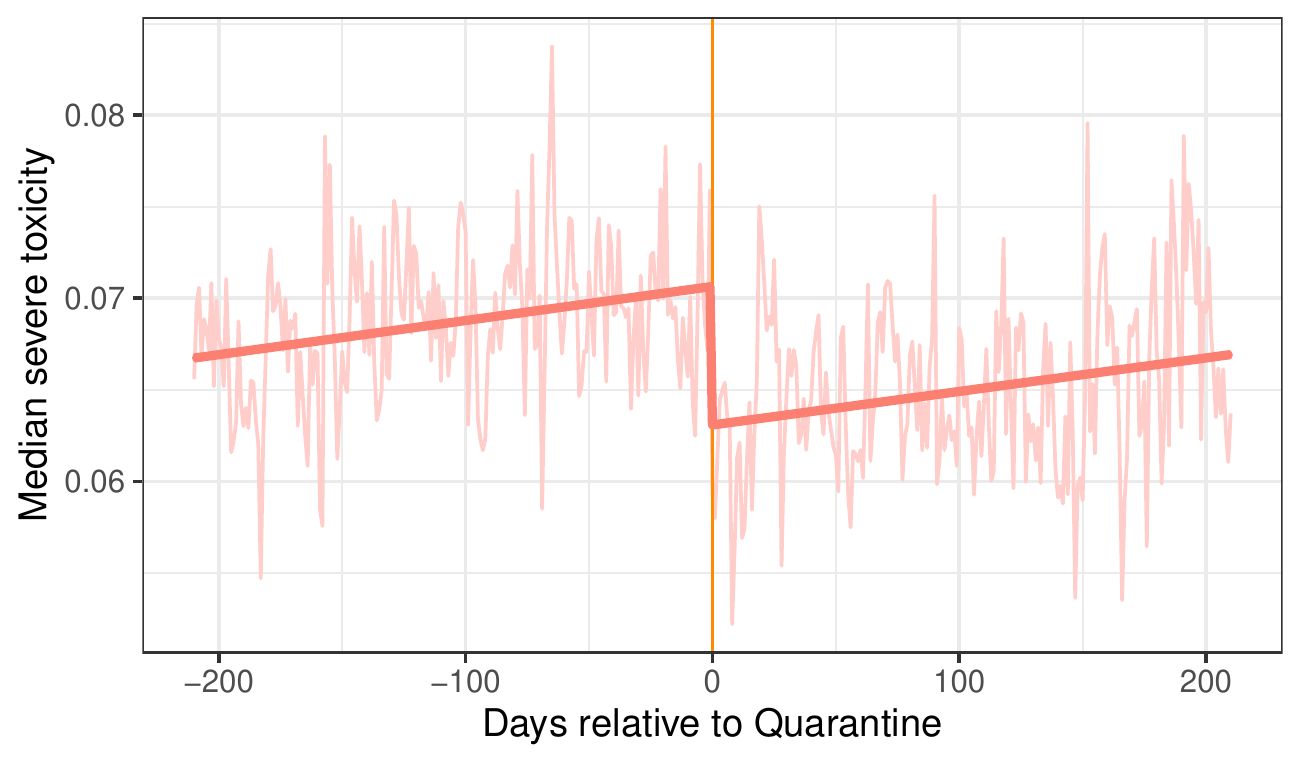}}
    \subcaptionbox{Fraction of severely toxic comments.\label{fig:ctd_itd_q_sevtox_prop_at_0_5}}
        {\includegraphics[width=0.495\textwidth]{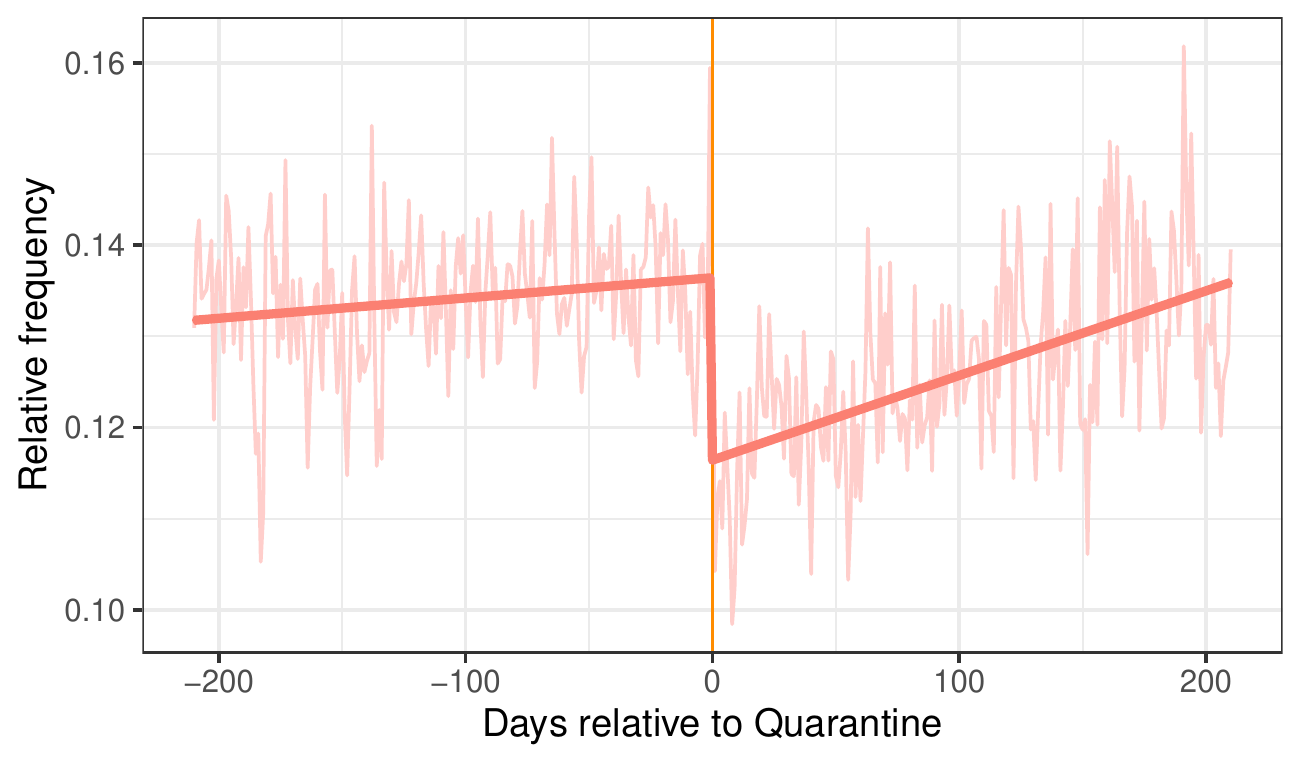}}
    \caption{ITS regressions of core users toxicity within \subr{The\_Donald} around the quarantine.}
    \Description{Both the median and fraction of severe toxicity show a noticeable drop upon quarantine, but due to an increasing trend afterward, toxicity levels slightly surpass pre-intervention levels at the end of a period of 210 days.}
    \label{fig:rq1_q_toxicity}
\end{figure}

We collected toxicity scores for 3.29M comments made by core users within \subr{The\_Donald} around the quarantine, which represent 32.1\% of the non-erased comment total for the same time span. Based on the ITS analyses shown in Figure~\ref{fig:rq1_q_toxicity}, during pre-quarantine the median severe toxicity score of core users' comments had a noticeable increasing linear trend. However, upon quarantine there was a significant immediate downward effect, with the median reaching its lowest value a few days later. Despite this drop in toxicity, the previous increasing trend reprised during post-quarantine, reaching median levels similar to those in pre-quarantine. Regarding ``severely toxic'' comments, there is a less marked increasing linear trend compared to the median, although both the immediate effect and the subsequent increase post-quarantine are much more evident. Additional analyses on the robustness of our results confirmed that these effects hold at different thresholds of severe toxicity used for labeling ``severely toxic'' comments, as well as with other Perspective toxicity scores (e.g., toxicity, insult, threat). According to the BSTS results reported in Table~\ref{tab:rq1_toxicity_bsts}, the causal effects can be considered statistically significant (posterior tail-area probability $pp\approx.001$), but as previously stated, there is a noticeable decay of the effects, with the relative frequency of comments classified as ``severely toxic'' reaching even higher levels compared to pre-quarantine. The quarantine thus had the strong immediate effect of reducing toxicity, but also the strong long-term effect of increasing it.

\begin{table}[t]
    \centering
    \caption{BSTS results of quarantine effects on core users toxicity within \subr{The\_Donald}. The lower the posterior tail-area probability (\textit{pp}), the higher the probability of a causal effect.}
    \label{tab:rq1_toxicity_bsts}
    \begin{tabular}{lrrcS[retain-explicit-plus]cc}
    \toprule
    & \multicolumn{3}{c}{Post-intervention avg. value} & \multicolumn{2}{c}{Relative effect (\%)} & \\
    \cmidrule(lr){2-4}\cmidrule(lr){5-6}
                                    & Actual  & Predicted & 95\% CI   & \multicolumn{1}{c}{\% Change} & 95\% CI & \textit{pp}\\ 
    \hline
    Median severe toxicity & .065 & .069 & [.068, .070] & -6 & [-7.9, -4] & .001\\
    Severely toxic comments     & .126 & .136 & [.132, .137]  & -6.4 & [-8.2, -4.8]  & .001\\
    \bottomrule
    \end{tabular}
\end{table}

\subsection{RQ2: Effects on the Quality of Shared News Articles within r/The\_Donald}
\label{sec:rq2-results}

\subsubsection{Political bias}
\label{sec:rq2-political-bias}
\begin{figure}[t]
    \subcaptionbox{General bias.\label{fig:q_bias}}
        {\includegraphics[width=.975\textwidth]{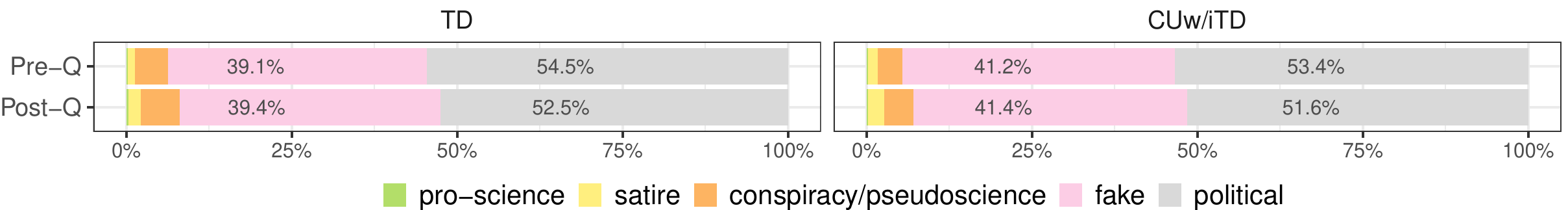}}
    \subcaptionbox{Political polarization.\label{fig:q_pol_bias}}
        {\includegraphics[width=.975\textwidth]{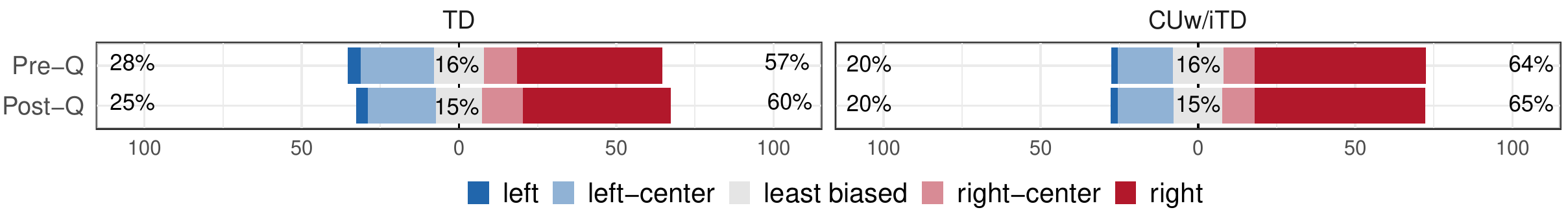}}
    \subcaptionbox{Factual reporting.\label{fig:q_factuality}}
        {\includegraphics[width=.975\textwidth]{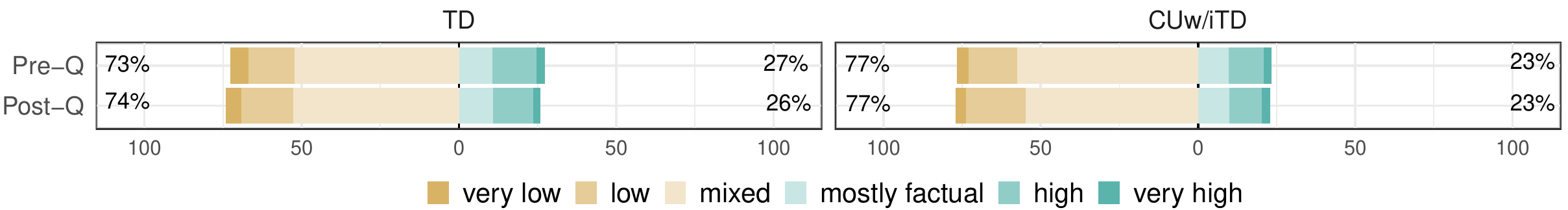}}
    \caption{Quality of news outlets in \subr{The\_Donald} around quarantine for all users (\dsname{TD}) and core users (\dsname{CUw/iTD}).}
    \Description{Upon quarantine, there is no significant change in terms of general bias, political polarization, and factual reporting of submitted news outlets.}
    \label{fig:rq2_q_mbfc}
\end{figure}

Most news articles (53\%) shared within \subr{The\_Donald} came from politically biased sources, as shown in Figure~\ref{fig:q_bias}, with the majority of these (62\%) falling at the right of the US political spectrum, as shown in Figure~\ref{fig:q_pol_bias}. 
News shared by core users were more biased to the right, compared to all users, with 64\% and 58\% respectively. After  quarantine, there was a small (3\%) but significant increase in polarization to the right for all users ($\chi^2 = 90.3; p <.001$), while for core users there was a non-significant increase of only 1\% ($\chi^2 = 0.641; p = .423$).

\begin{figure}[t]
    \subcaptionbox{Content activity (quarantine).\label{fig:ctd_otd_q_posts}}
        {\includegraphics[width=0.495\textwidth]{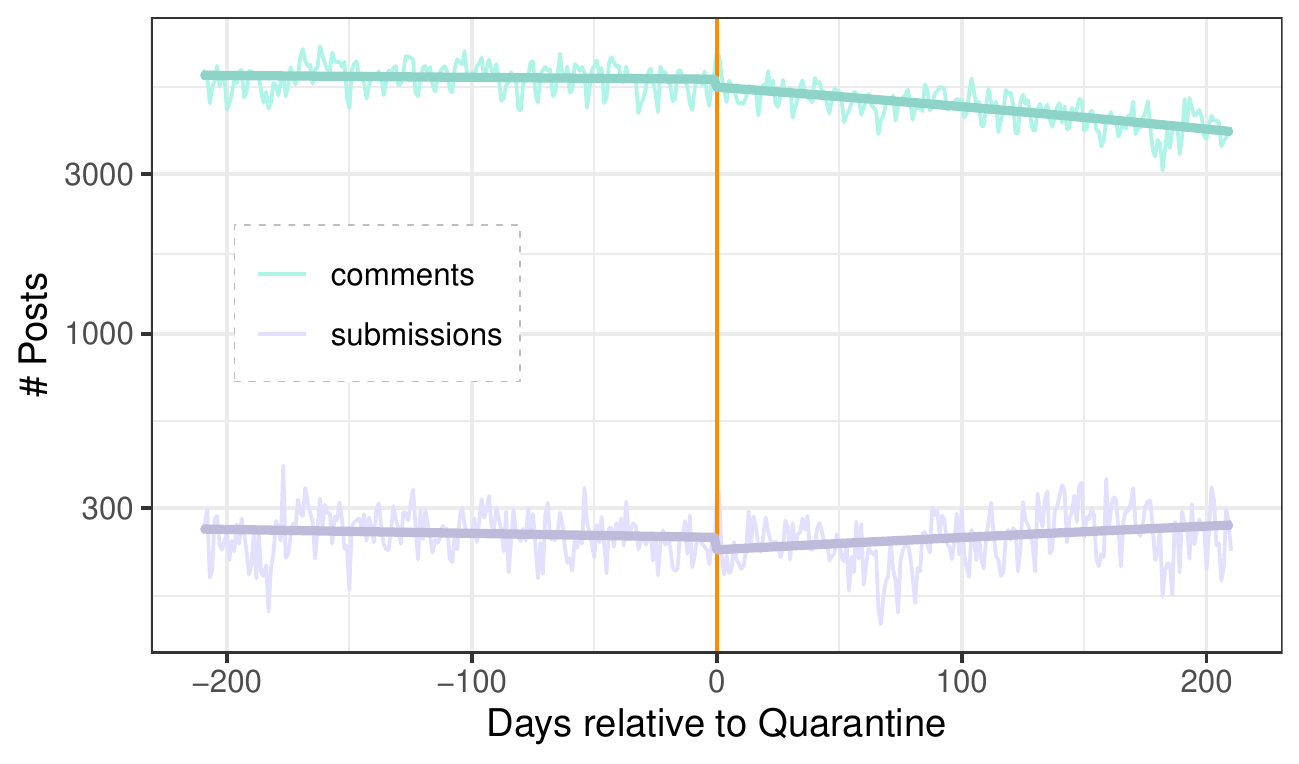}}
    \subcaptionbox{Content activity (restriction).\label{fig:ctd_otd_r_posts}}
        {\includegraphics[width=0.495\textwidth]{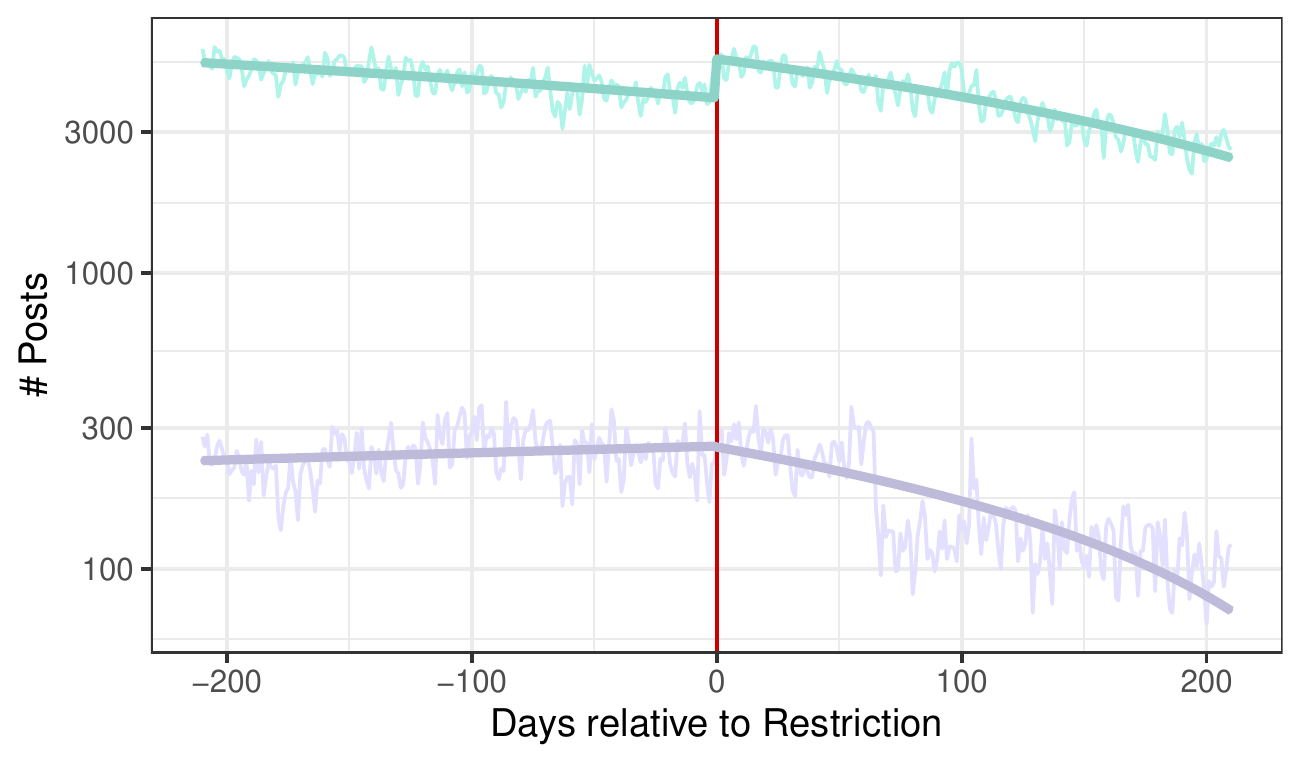}}
    \par\bigskip 
    \subcaptionbox{Daily active users (quarantine).\label{fig:ctd_otd_q_dau}}
        {\includegraphics[width=0.495\textwidth]{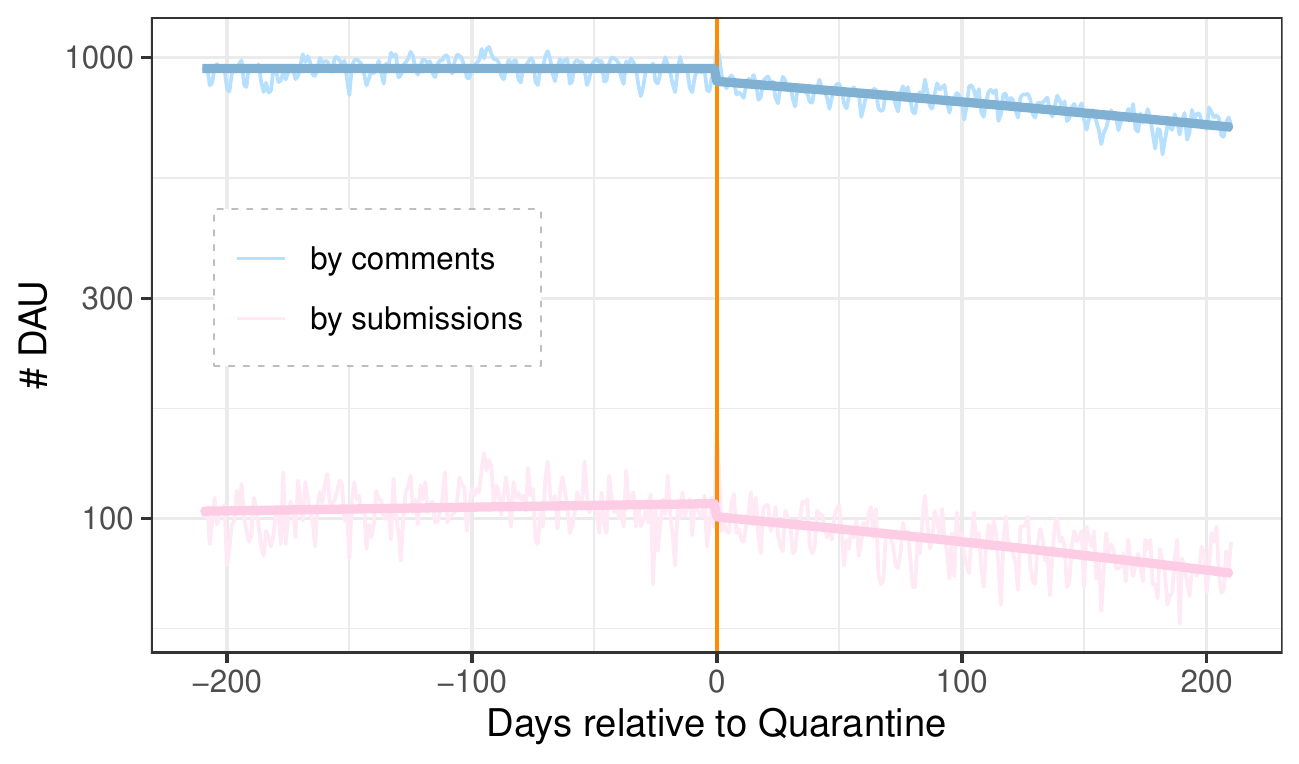}}
    \subcaptionbox{Daily active users (restriction).\label{fig:ctd_otd_r_dau}}
        {\includegraphics[width=0.495\textwidth]{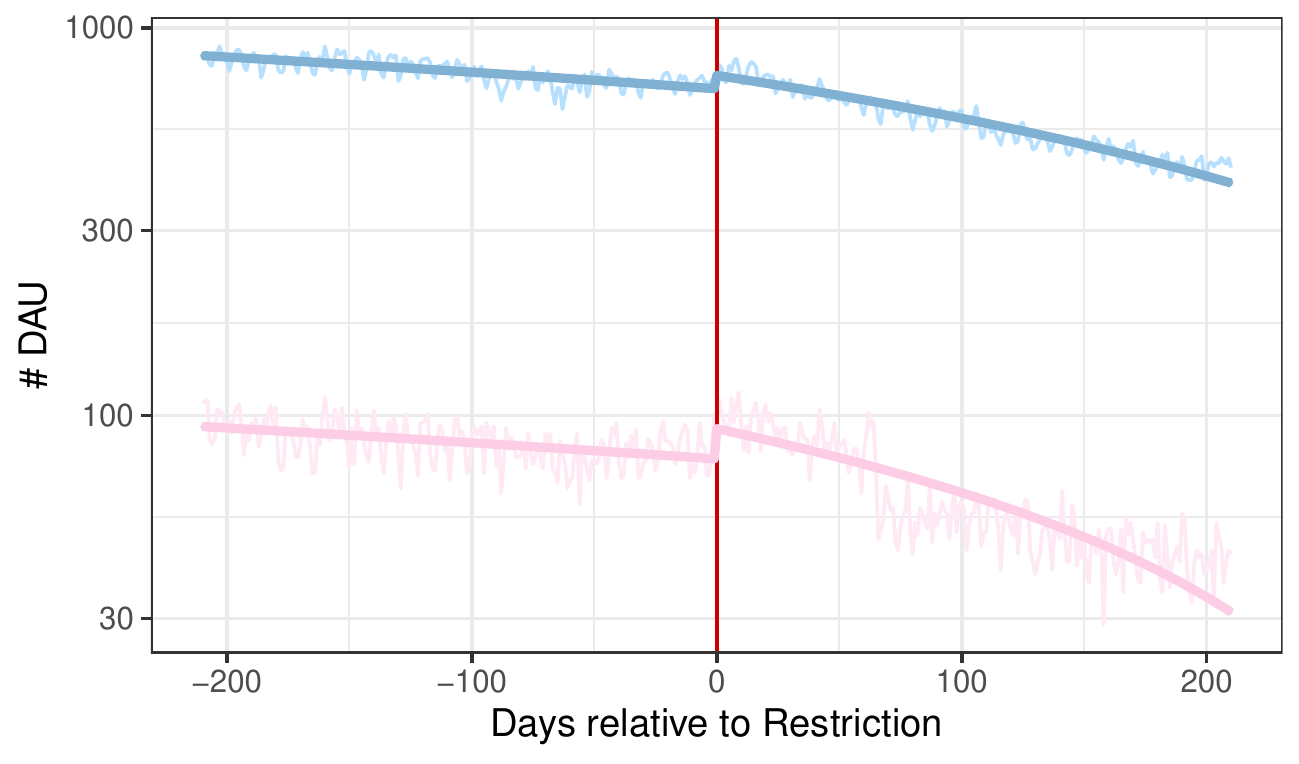}}
    \caption{ITS regressions of core users activity outside of \subr{The\_Donald} around quarantine and restriction.}
    \Description{In general, there is moderate a decline in activity upon quarantine, except for the number of submissions, which increases moderately; upon restriction, all activity measures decline sharply.}
    \label{fig:rq3_activity}
\end{figure}

\subsubsection{Factual reporting}
\label{sec:rq2-factual-reporting}
Figure~\ref{fig:q_factuality} shows that around 75\% of shared news content came from outlets in the lower half of the factual reporting spectrum, with 40\% of the total shared articles deemed as mostly \emph{fake} (see definition in \S\ref{sec:political_polarization}). 
On average, core users shared slightly less reliable content compared to all users, but there was no significant change in factual reporting between pre- and post-quarantine ($\chi^2 = 1.0353; p=.308 $), whereas the factuality decrease for all users was small (-1.2\%) but significant ($\chi^2 = 47.37; p<.001$).

\subsection{RQ3: Effects of Multiple Interventions outside of r/The\_Donald}
\label{sec:rq3-results}

\subsubsection{Activity}
\label{sec:rq3-activity}
In general, core users activity outside of \subr{The\_Donald} suffered an immediate and sustained decrease in the post-quarantine period (consistent with activity within the subreddit), except for the number of daily submissions, which manifested an increasing trend post-intervention. This means that the remaining active core users increased their activity in terms of submissions to other subreddits, after a slight decrease upon quarantine. Indeed, the BSTS analysis reported in Table~\ref{tab:rq3_activity_bsts} indicates that there is no significant effect for submissions after quarantine (relative effect of -0.9\%, $pp = .343$), whilst for the other metrics we measured significant effects ($pp=.001$), with relative effects ranging from -14.5\% to -17.9\%.

The situation changes in the post-restriction period. On the one hand, even if the number of DAU that commented significantly decreased (relative effect of -20.6\%, $pp=.001$), the number of comments had an immediate slight increase upon restriction and a less pronounced decreasing trend (relative effect of -6.6\%, $pp=0.051$), as illustrated in Figures~\ref{fig:ctd_otd_r_posts} and~\ref{fig:ctd_otd_r_dau}. On the other hand, the restriction had a strong and significant effect ($pp=.001$) on both the number of submissions (-33.2\%) and submission DAU (-24.6\%). In part, this decrease is likely due to the migration of members of \subr{The\_Donald} to a different platform ({\small\texttt{thedonald.win}}) upon restriction. 

\begin{table}[t]
    \centering
    \caption{BSTS results of quarantine (Q) and restriction (R) effects on core users activity outside \subr{The\_Donald}. The lower the posterior tail-area probability (\textit{pp}), the higher the probability of a causal effect.}
    \label{tab:rq3_activity_bsts}
    \begin{tabular}{lrrcScc}
    \toprule
    & \multicolumn{3}{c}{Post-intervention avg. value} & \multicolumn{2}{c}{Relative effect (\%)} & \\
    \cmidrule(lr){2-4}\cmidrule(lr){5-6}
                                    & Actual  & Predicted & 95\% CI   & \multicolumn{1}{c}{\% Change} & 95\% CI & \textit{pp}\\ 
    \hline
    Q submissions     & 246 & 249 & [237, 260] & -0.9 & [-5.4, 3.7] & .343\\
    Q comments        & 4748 & 5780 & [5610, 5948]  & -17.9 & [-21, -15]  & .001\\
    Q submission {\small DAU}  & 88 & 106 & [102, 109] & -16.2 & [-20, -13] & .001\\
    Q comment {\small DAU}     & 797 & 932 & [909, 950]  & -14.5 & [-16, -12]  & .001\\
    R submissions     & 165 & 247 & [234, 263] & -33.2 & [-39, -28] & .001\\
    R comments        & 3886 & 4161 & [3835, 4472]  & -6.6 & [-14, 1.2]  & .051\\
    R submission {\small DAU}  & 62 & 82 & [78, 86] & -24.6 & [-30, -19] & .001\\
    R comment {\small DAU}     & 575 & 725 & [7176, 7781]  & -20.6 & [-26, -15]  & .001\\
    \bottomrule
    \end{tabular}
\end{table}

\subsubsection{Toxicity}
\label{sec:rq3-toxicity}
\begin{figure}[t]
    \subcaptionbox{Median severe toxicity scores (quarantine).\label{fig:ctd_otd_q_sevtox_median}}
        {\includegraphics[width=0.495\textwidth]{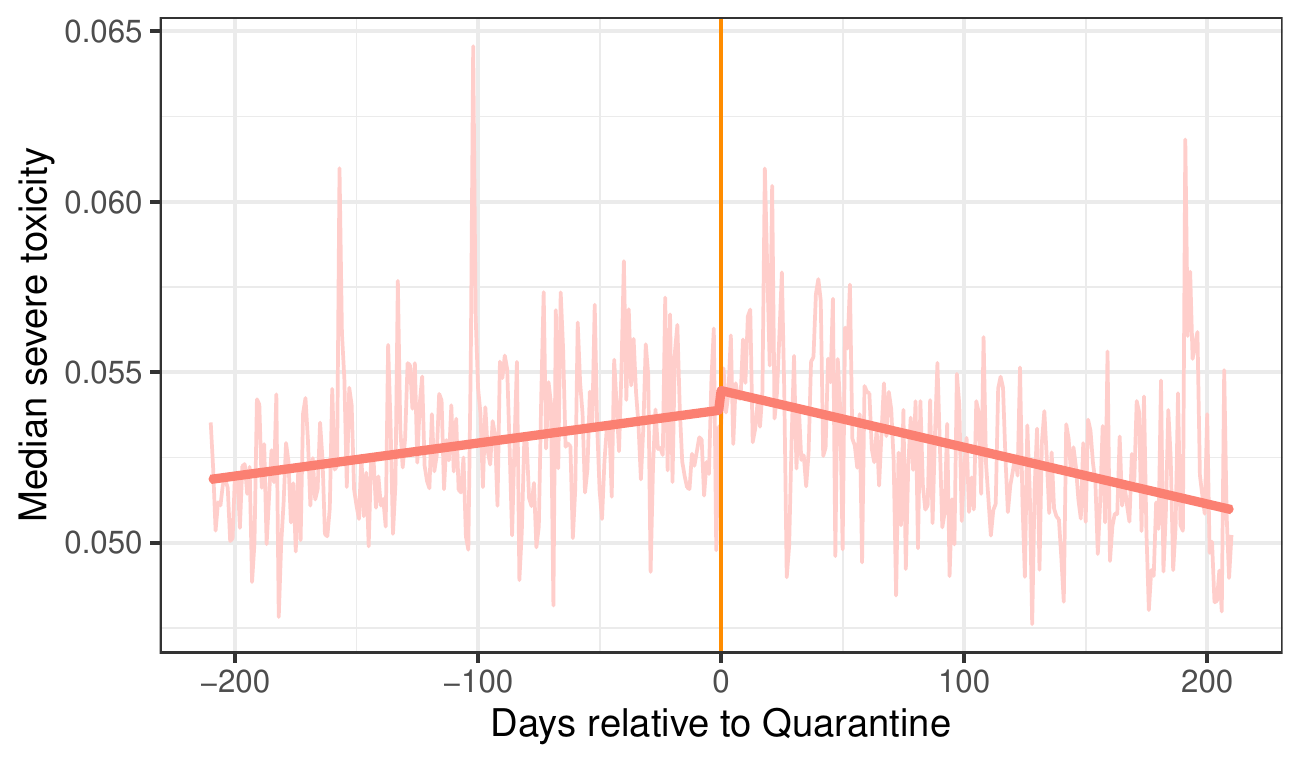}}
    \subcaptionbox{Median severe toxicity scores (restriction).\label{fig:ctd_otd_nr_sevtox_median}}
        {\includegraphics[width=0.495\textwidth]{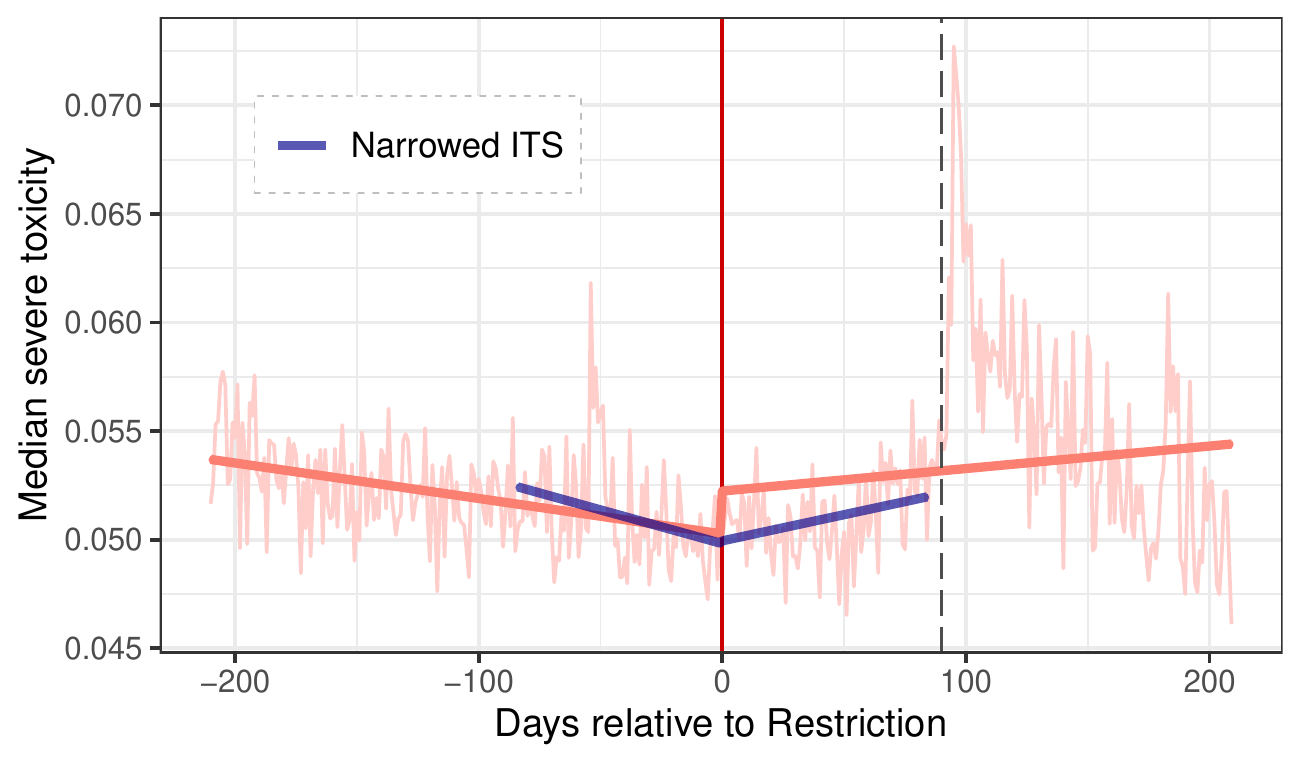}}
    \par\bigskip 
    \subcaptionbox{Fraction of severely toxic comments (quarantine).\label{fig:ctd_otd_q_sevtox_prop_at_0_5}}
        {\includegraphics[width=0.495\textwidth]{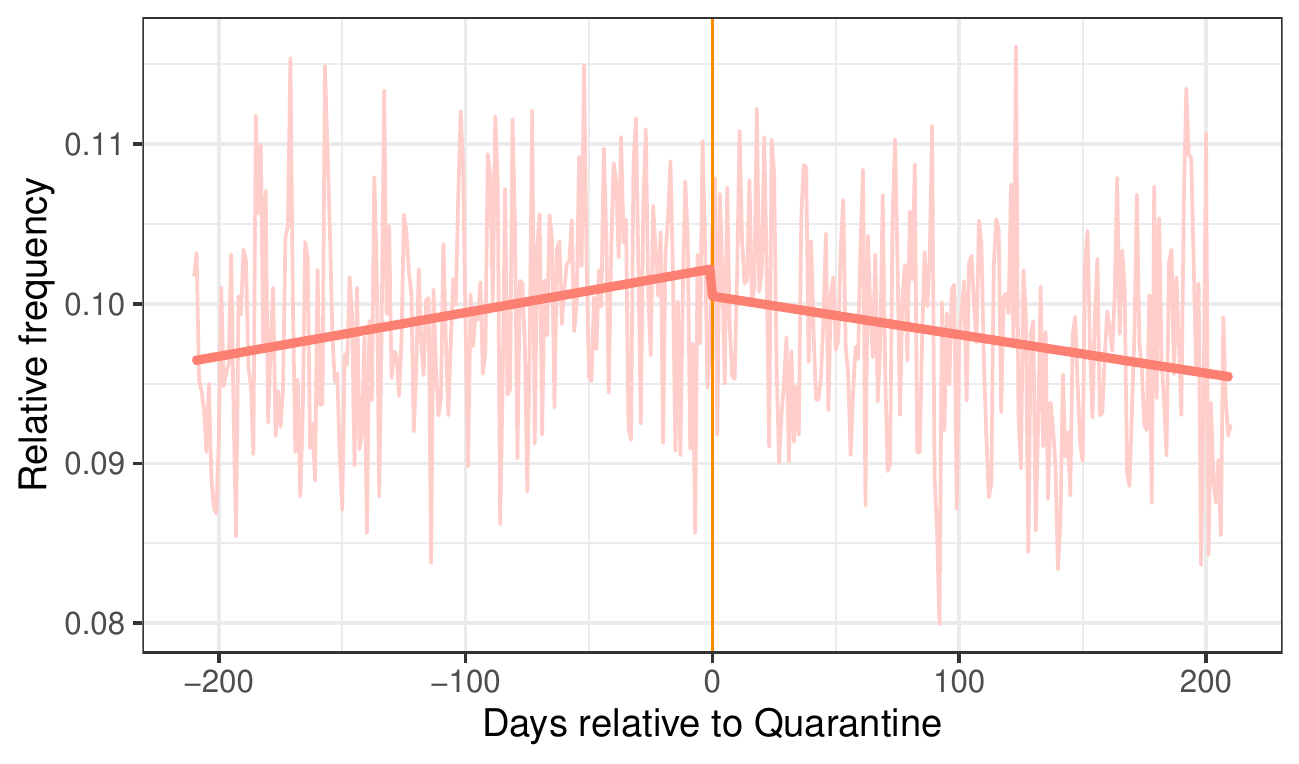}}
    \subcaptionbox{Fraction of severely toxic comments (restriction).\label{fig:ctd_otd_nr_sevtox_prop_at_0_5}}
        {\includegraphics[width=0.495\textwidth]{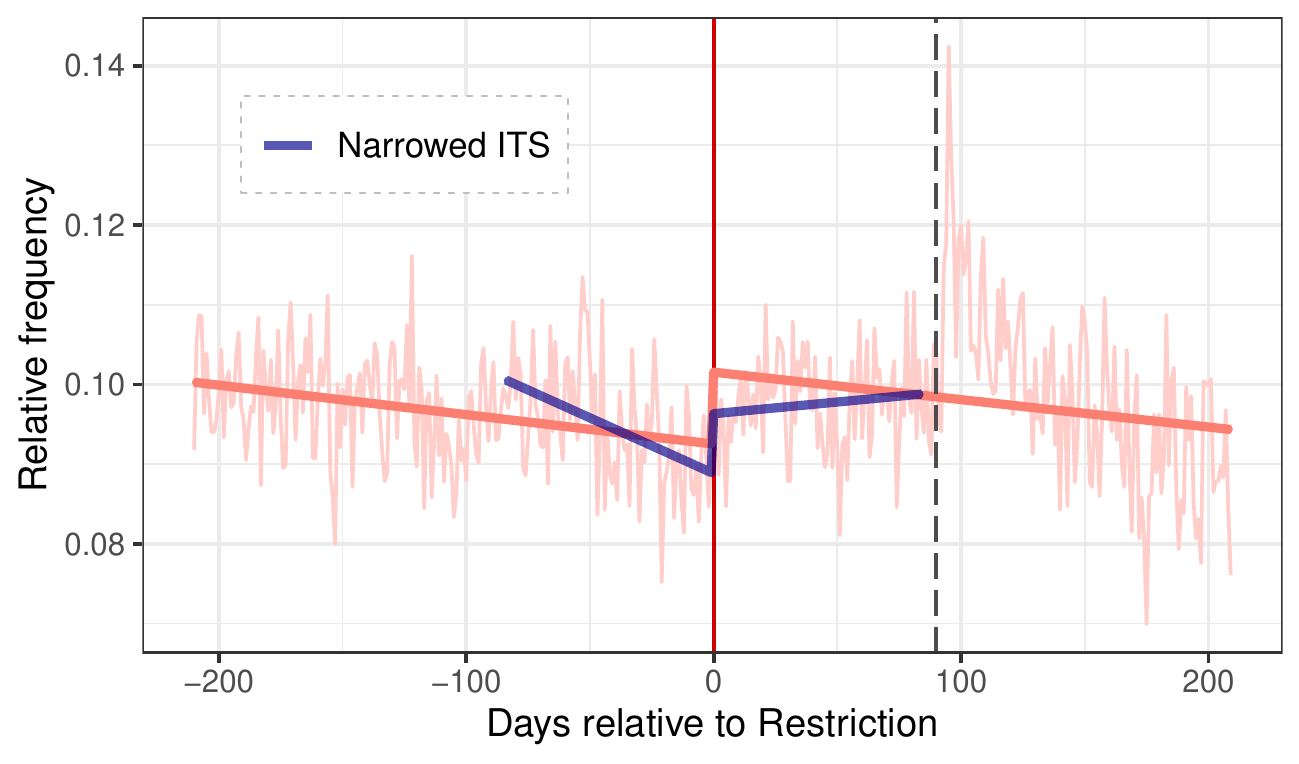}}
    \caption{ITS regressions of core users toxicity outside of \subr{The\_Donald} around the quarantine and the restriction. A vertical dashed line indicates the start of the George Floyd protests, after which toxicity surged. A narrowed (±12 weeks) ITS regression analysis (blue-colored) excludes the protests.}
    \Description{Upon quarantine, there is a decreasing trend in toxicity that contrasts with an increasing trend pre-quarantined. Upon restriction, there is a remarkable surge in toxicity due to the George Floyd protests. After narrowing the data to exclude the protests, upon restriction a decreasing trend in toxicity changes direction and becomes increasing.}
    \label{fig:rq3_i_toxicity}
\end{figure}

We collected toxicity scores for 3.19M comments made by core users outside of \subr{The\_Donald}, which is a lower number with respect to the 3.5M comments made within \subr{The\_Donald} by the same users during the same time frame. Regarding the quarantine, the pre-intervention periods of both median values and relative frequency of severely toxic comments had a moderate increasing linear trend, which upon quarantine became decreasing, as shown in Figures~\ref{fig:ctd_otd_q_sevtox_median} and~\ref{fig:ctd_otd_q_sevtox_prop_at_0_5}. The BSTS results summarized in Table~\ref{tab:rq3_toxicity_bsts} show that the effect for the median values was not significant ($pp=.125$), while it is significant for severely toxic comments ($pp=.021$). 

Concerning the restriction, the situation is more complex, as there is a remarkable surge in toxicity during the weeks of the George Floyd protests, started on May 26, 2020. 
These protests had a marked polarizing effect in the US public opinion, particularly between liberals and conservatives~\cite{reny2021opinion}. 
Our result is consistent with the difficulties encountered by~\citeauthor{horta2021platform} in analyzing toxicity in \subr{The\_Donald}. 
For these reasons, we performed additional analyses with narrowed pre-post restriction periods of ±12 weeks, thus excluding these protests.
In the pre-restriction span, the median values and relative frequencies showed a decreasing linear trend, clearly visible in Figures~\ref{fig:ctd_otd_nr_sevtox_median} and~\ref{fig:ctd_otd_nr_sevtox_prop_at_0_5}. All of the post-restriction trends show an upward trend, except for the 30-week period pertaining to relative frequency of toxic comments in which is downward. Based on the BSTS analysis, whose results are in Table~\ref{tab:rq3_toxicity_bsts}, post-intervention effects are significant ($pp \leq 0.003$), except for the median values during quarantine ($pp=.125$) and narrowed restriction ($pp=.462$). 

\begin{table}[t]
    \centering
    \caption{BSTS results of quarantine (Q) and restriction (R) effects on core users toxicity outside \subr{The\_Donald}. Pre-post periods span 30 weeks, except those narrowed (n.) to 12 weeks to exclude the George Floyd's protests. The lower the posterior tail-area probability (\textit{pp}), the higher the probability of a causal effect.}
    \label{tab:rq3_toxicity_bsts}
    \begin{tabular}{lrrcS[retain-explicit-plus]cc}
    \toprule
    & \multicolumn{3}{c}{Post-intervention. avg. value} & \multicolumn{2}{c}{Relative effect (\%)} & \\
    \cmidrule(lr){2-4}\cmidrule(lr){5-6}
                                    & Actual  & Pred. & 95\% CI   & \multicolumn{1}{c}{\% Change} & 95\% CI & \textit{pp}\\ 
    \hline
    Q Median severe toxicity      & .052 & .053 & [.052, .054] & -0.73 & [-2, 0.45] & .125\\
    Q Severely toxic comments  & .098 & .100 & [.098, .102]  & -2.1 & [-4.3, -0.13]  & .021\\
    R Median severe toxicity     & .053 & .051 & [.050, .052] & +3.9 & [2.5, 5.6] & .001\\
    R Severely toxic comments & .098 & .095 & [.093, .097]  & +3.3 & [1.3, 5.7]  & .003\\
    R (n.) Median severe toxicity & .053 & .051 & [.050, .052] & -0.14 & [-1.7, 1.4] & .462\\
    R (n.) Severely toxic comments & .098 & .094 & [.092, .097]  & +3.5 & [0.8, 6.1]  & .004\\
    \bottomrule
    \end{tabular}
\end{table}

\subsubsection{Political bias}
\label{sec:rq3-political-bias}
\begin{figure}[t]
    \subcaptionbox{General bias.\label{fig:ctd_otd_i_bias}}
        {\includegraphics[width=0.75\textwidth]{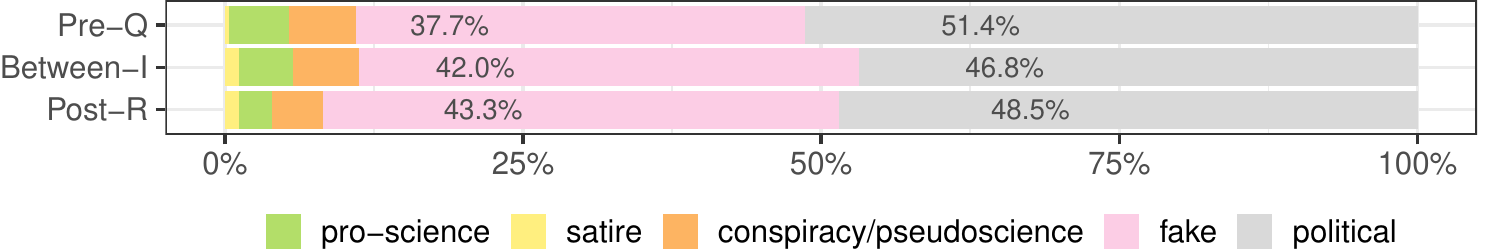}}
    \par\bigskip 
    \subcaptionbox{Political polarization.\label{fig:ctd_otd_i_pol_bias}}
        {\includegraphics[width=0.75\textwidth]{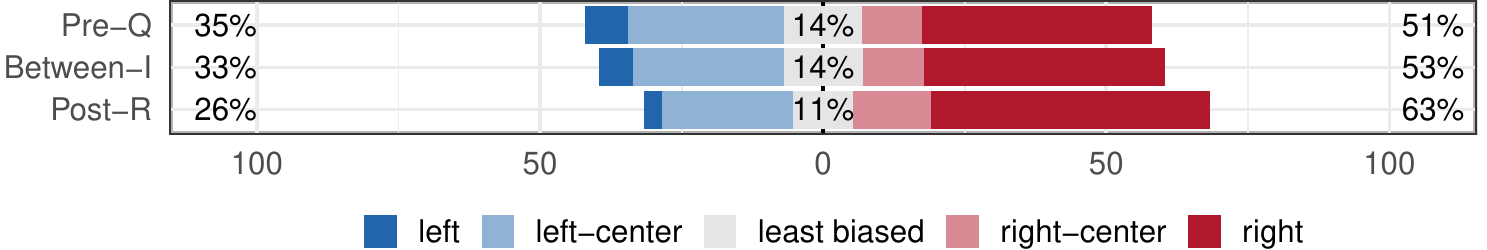}}
    \par\bigskip 
    \subcaptionbox{Factual reporting.\label{fig:ctd_otd_i_factuality}}
        {\includegraphics[width=0.75\textwidth]{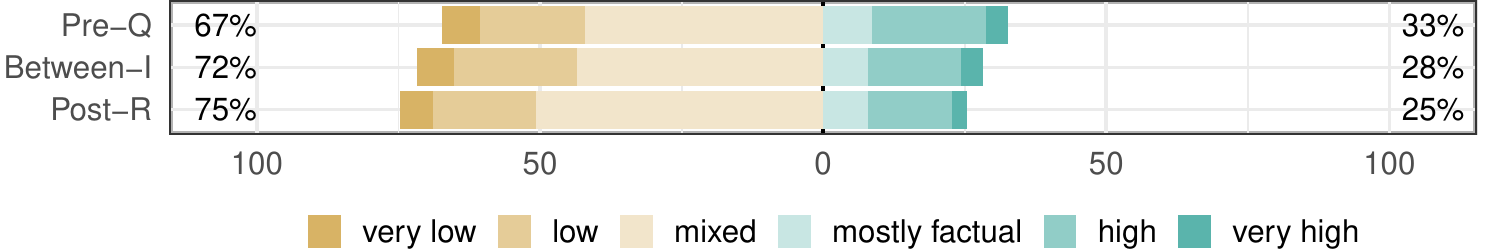}}
    \caption{Quarantine (Q) and restriction (R) effects on the quality of news articles shared outside of \subr{The\_Donald} by core users.}
    \Description{Upon each intervention the political polarization increases to the right and the factual reporting of shared news decreases visibly.}
    \label{fig:rq3_i_mbfc}
\end{figure}

During the time frame of the study, there were 399K submissions made by core users outside of \subr{The\_Donald}, of which 91K have an external link (22.9\%), and 49K are present in MBFC (12.37\% of the submission total). Around half of the shared news links pertained to politically biased outlets. Before any intervention, core users shared fewer articles from outlets biased to the right of the political spectrum outside of \subr{The\_Donald} (51\%) compared to within (64\%). However, upon both quarantine and restriction there was a significant increase in bias to the right in content posted in other subreddits, respectively reaching 53\% ($\chi^2 = 13.07; p <.001$) and 63\% ($\chi^2 = 95.8; p <.001$) in the post-interventions periods, as shown in Figure~\ref{fig:ctd_otd_i_pol_bias}.

\subsubsection{Factual reporting}
\label{sec:rq3-factual-reporting}
We noticed a similar phenomenon in the decrease of factuality of shared news after each intervention, as it can be seen in Figure~\ref{fig:rq3_i_mbfc}. Moreover, before any intervention, core users shared more factual content outside of \subr{The\_Donald} compared to within, as visible in Figures~\ref{fig:q_factuality} and~\ref{fig:ctd_otd_i_factuality}. For instance, in the pre-quarantine period the share of news articles from unreliable sources was 67\% outside of \subr{The\_Donald} and 73\% within. Between interventions, the low-factuality outside of \subr{The\_Donald} increased to 72\% ($\chi^2 = 90.1; p < .001$), and in the post-restriction period increased to 75\% ($\chi^2 = 28.8; p < .001$). This change is most likely related to the increase in proportion of articles from \emph{fake} outlets after each restriction, as shown in Figure~\ref{fig:ctd_otd_i_bias}, which during pre-quarantine period was 37.7\%, then 42\% between the restrictions, and finally 43.3\% in the post-restriction period. 

\begin{table}[t]
    \centering
    \caption{Rank-biased overlap (RBO) between group community proclivity (top 50 subreddits) by content kind and paired intervention periods. The higher the RBO scores, the more similar the proclivity between periods.}
    \label{tab:rbo}
    \begin{tabular}{lrr}
    \toprule
    & \multicolumn{2}{c}{Proclivity RBO}\\
    \cmidrule(lr){2-3}
    & Submissions & Comments \\
    \hline
    Pre-Q \ampsep{} Between-I  & .329   &  .658 \\
    Between-I \ampsep{} Post-R & .299   &  .431 \\
    Pre-Q \ampsep{} Post-R     & .145   &  .354 \\
    \bottomrule
    \end{tabular}
\end{table}

\begin{figure*}[t]
    \centering
    \includegraphics[width=1\textwidth]{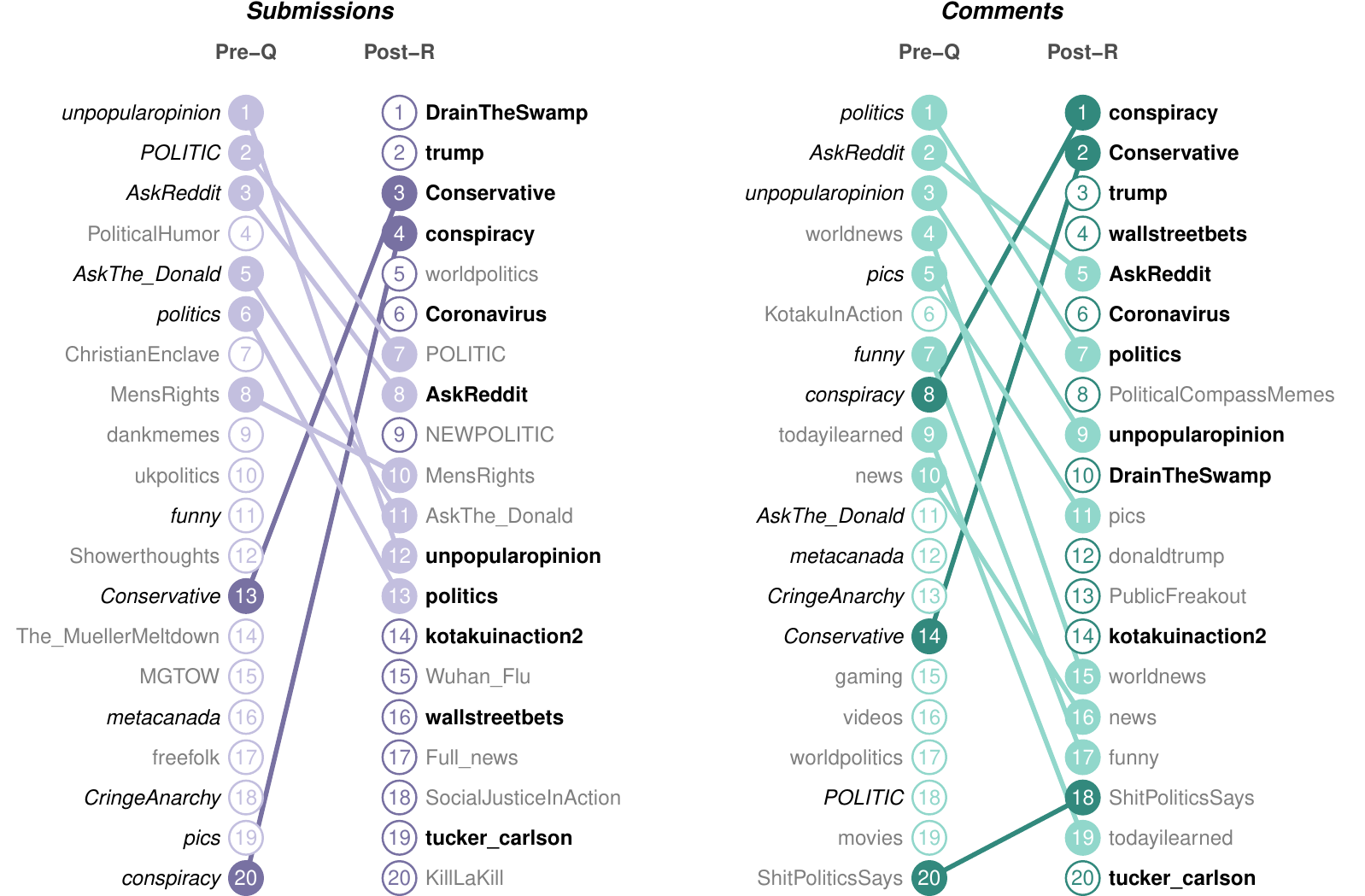}
    \caption{Top 20 subreddits ranked by group community proclivity. Subreddits in italic and bold are present in both content lists in Pre-Q and Post-R, respectively. Dark-colored subreddits increased their ranking after the intervention, while lightly-colored subreddits decreased their ranking after the intervention.}
    \Description{Between the pre-quarantine and post-restriction periods, the group community proclivity for submissions changed more than that of comments, with many newcomer subreddits in the top 20 for the former. However, in the case of both content kinds there were two subreddits that climbed to the top or near-top positions: conspiracy and conservative.}
    \label{fig:proclivity-ranking}
\end{figure*}

\subsubsection{Group community proclivity}
\label{sec:rq3-platform-effects}
Results of intervention effects on group community proclivity denote a different proclivity dynamic for submissions and comments, as reported in Table~\ref{tab:rbo}. Specifically, the rankings of most prominent subreddits based on submissions are less similar among intervention periods compared to those based on comments. In particular, the change in similarity for Pre-Q and Between-I by submissions (.329) is much lower than the respective value by comments (.658). Thus, after the quarantine, core users showed a higher proclivity for a much different set of subreddits in which to post submissions, compared to comments. This phenomenon can also be seen in the changes in proclivity between Pre-Q and Post-R periods depicted in Figure~\ref{fig:proclivity-ranking}, in which the subreddits that are present in the top 20 lists for both periods, are fewer for the lists based on submissions than for those based on comments. Incidentally, three subreddits climbed to the top-4 positions in both content types during Post-R: \subr{conspiracy}, \subr{Conservative}, \subr{trump}.

\begin{figure}[t]
    \subcaptionbox{Participation to distinct subreddits (quarantine).\label{fig:ctd_q_subreddits_its}}
        {\includegraphics[width=0.495\textwidth]{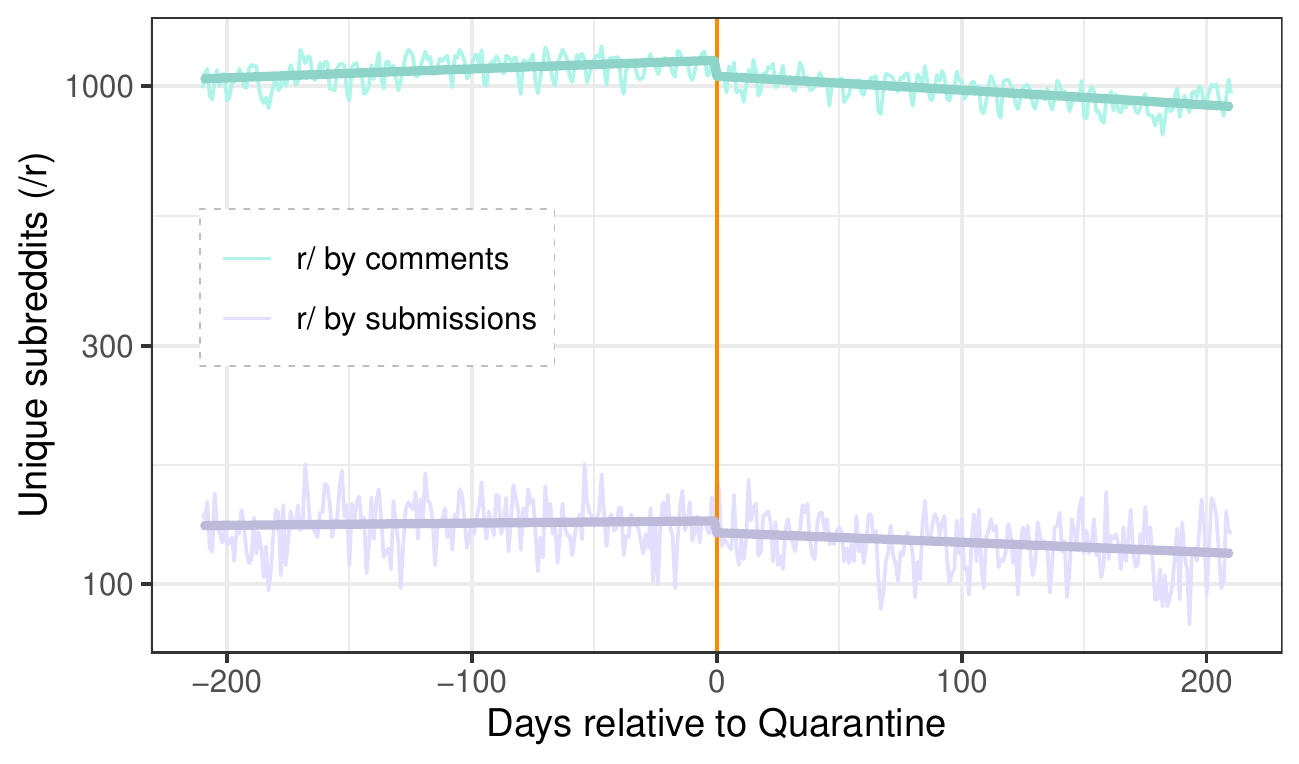}}
    \subcaptionbox{Participation to distinct subreddits (restriction).\label{fig:ctd_r_subreddits_its}}
        {\includegraphics[width=0.495\textwidth]{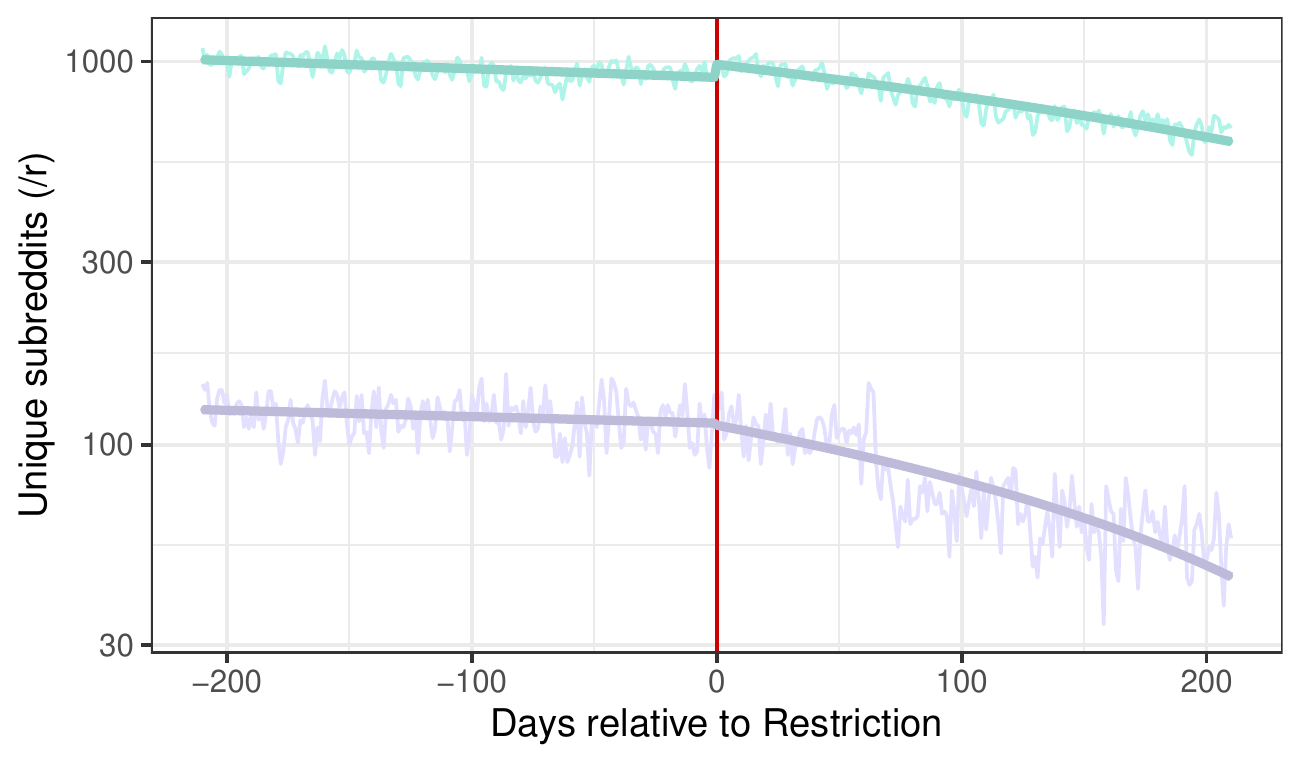}}
    \caption{ITS regressions of core users participation (either by commenting or by submitting content) to distinct subreddits around the quarantine and the restriction.}
    \Description{Upon quarantine, there is a very weak decreasing trend in the unique subreddits in which cor users participate. However, upon the restriction, the decreasing trend is more evident for comments and remarkable for submissions.}
    \label{fig:rq3_unique_subreddits}
\end{figure}

It should be noted that these changes in proclivity could have been influenced by changes in the number of subreddits in which core users participated. Hence, we also investigated the effect of interventions on the number of daily distinct subreddits. Before quarantine, the daily average of distinct subreddits in which core users of \subr{The\_Donald} participated ---as a whole--- was of 132.3 by submissions and 1078.9 by comments. Upon quarantine, our BSTS analysis indicates a significant ($pp=.001$) decrease of -8.6\% by submissions and -11\% by comments, albeit both immediate effects and trends are moderately downward, as visible from the ITS regression in Figure~\ref{fig:ctd_q_subreddits_its}. Upon restriction, however, the post-intervention downward trend and significant ($pp=.001$) relative effects are more pronounced, especially in the case of submissions at -33\% with respect to comments at -11\%, as shown in Figure~\ref{fig:ctd_r_subreddits_its}. Comparing results from Figure~\ref{fig:ctd_q_subreddits_its} with those of Figures~\ref{fig:rq1_q_activity} and~\ref{fig:rq3_activity} also allows gaining insights into the tactics ---and their efficacy--- used by core users of \subr{The\_Donald} to circumvent Reddit’s interventions. After quarantine, core users posted many more submissions outside of \subr{The\_Donald}, as a way to elude the intervention. This behavior resulted in a markedly different submissions proclivity. However, the new submissions did not attract much attention since they received few comments, as testified by a relatively stable comment-based proclivity.

%% file: discussion.tex
\section{Discussion}
\label{sec:discussion}

Our results allow to evaluate the effectiveness of quarantining and restricting \subr{The\_Donald}, both within the subreddit as well as in other parts of Reddit. In summary, the quarantine had mild positive effects with respect to reducing user activity and strong positive immediate effects for reducing toxicity within \subr{The\_Donald}. These effects were stronger for core users of \subr{The\_Donald} with respect to the other users. On the contrary however, quarantining \subr{The\_Donald} had no significant effects on news quality and, more worryingly, had the strong negative long-term effect of increasing toxicity. In particular, the degree of toxicity within \subr{The\_Donald} reverted to, and even surpassed, pre-intervention levels around 6 months after the quarantine.

Regarding the effects that quarantining and restricting \subr{The\_Donald} had on other subreddits, we found positive effects towards reducing user activity. Specifically, while the quarantine produced mild effects, the restriction was instead very effective at reducing activity. However, both the quarantine and restriction caused an overall increase in toxicity. In detail, the quarantine caused a small immediate increase and strong long-term decrease in toxicity, while the restriction caused moderate immediate increase as well as strong long-term increase in toxicity. Moreover, both interventions also had marked negative effects on the quality of the shared news, which gradually became more polarized and less factual.

Overall, our results support the following conclusions:
\renewcommand\labelenumi{\bfseries{\theenumi.}}
\begin{itemize}
    \item The restriction produced stronger effects platform-wise than the quarantine.
    \item Core users suffered stronger effects than other users.
    \item Both interventions did well at reducing activity. 
    \item At the same time however, both interventions produced an increase in toxicity.
    \item Both also caused affected users to share more polarized and less factual news.
\end{itemize}

In the next paragraphs we discuss our results in the context of the current state of knowledge, the main limitations of our work, and we identify possible directions for future research.

\subsection{Temporal Variations of Effects}
\label{sec:discussion-long-term}
Longitudinal studies on moderation intervention effects are almost absent in literature, with most existing studies making no distinction between short-term and long-term effects. Exceptions are \citeauthor{chandrasekharan2020quarantined} who raised attention to the importance of verifying «whether the subreddits that come out of quarantine maintain improved discourse over long periods of time or whether they return to incendiary behavior», although without providing results~\cite{chandrasekharan2020quarantined}. Instead, \citeauthor{jhaver2021evaluating} evaluated the effects of deplatforming influential Twitter users and claimed that this had long-term positive consequences~\cite{jhaver2021evaluating}. In this regard our results suggest caution, as we found important differences between the short-term and long-term effects of quarantine and restriction. 
Such differences were striking for toxicity within \subr{The\_Donald} after the quarantine. In that case, the strong initial decrease in toxicity started decaying a few days after the intervention, up to the point that 6 months later, toxicity had surpassed pre-intervention levels.

These results relate to the existing literature in a number of ways. Firstly, \citeauthor{chandrasekharan2020quarantined} pointed that particular attention should be devoted to monitoring toxic behavior when quarantines are lifted~\cite{chandrasekharan2020quarantined}, implying that removing an intervention might revert the positive effect that it originally had. Here however, we observed a progressive reversal of the effect just a few days after the quarantine was enforced (not lifted). Our results thus go beyond \citeauthor{chandrasekharan2020quarantined}'s concerns, suggesting that intervention effects should be continuously evaluated through time, not only in the aftermath of major changes. In addition, our results about the temporal variations of intervention effects also suggest that more attention should be paid to this aspect. Therefore, in future studies it would be interesting to separately evaluate short- and long-term effects. Moreover, results of previous studies that do not make this distinction should be taken with care.

Our results also provide useful guidance for policing online platforms, in that platform administrators should continuously monitor the effects of their interventions and, in case, should reiterate them when the initial advantages have decayed. Some interventions should thus not only be seen as a one-off medicine, but rather as continuous or recurring treatments, depending on the need.


\subsection{Spillover Effects}
\label{sec:discussion-spillover}
When a community suffers a moderation intervention, part of its members migrate to other communities or platforms. This long-studied behavior is inevitable when a community is permanently closed (e.g., deplatformed)~\cite{kumar2011understanding,newell2016user,chandrasekharan2017you,horta2021platform}. However, it has been shown that even milder interventions might induce some users to migrate~\cite{chandrasekharan2020quarantined}. When evaluating interventions it is thus important to assess changes not only within the moderated community, but also in other communities to which affected users might have migrated. We call the latter type of changes \textit{spillover effects}.

\citeauthor{chandrasekharan2020quarantined} saw such effects in the quarantines of \subr{The\_Donald} and \subr{TheRedPill}, finding that the intervention decreased the activity levels and influx of new users in many other communities frequented by users of \subr{The\_Donald}~\cite{chandrasekharan2020quarantined}. In other words, not only did the quarantine reduce activity and participation within the moderated subreddit, but it also had a similar effect in related subreddits. Our analysis confirms these results. Then, based on their findings about user activity, \citeauthor{chandrasekharan2020quarantined} concluded that «quarantining \subr{The\_Donald} did not spread the infection to other parts of Reddit»~\cite{chandrasekharan2020quarantined}. Our results on toxicity and quality of shared news provide additional evidence to evaluate this claim. In detail, we found that the quarantine caused a small immediate surge and a strong long-term reduction in the toxicity of core users of \subr{The\_Donald} when they commented in other subreddits. However, it also caused those users to share much more low-quality (i.e., politically biased and less factual) information. Overall, our results paint a more complex and nuanced picture of the effects that quarantining \subr{The\_Donald} had on other parts of Reddit. In conclusion, our results call for additional efforts at assessing spillover effects of moderation interventions, especially for milder interventions compared to deplatforming.


\subsection{Side Effects and Nuanced Intervention Effects}
\label{sec:discussion-side-effects}
Our work, as well as most existing studies on the effectiveness of Reddit's moderation interventions, investigated quarantines, restrictions, and bans~\cite{chandrasekharan2017you,saleem2018aftermath,chandrasekharan2020quarantined,horta2021platform}. All three have different goals: quarantines deter interaction, restrictions limit content creation, and bans permanently shut subreddits. Still, it is worth noting that while their mechanics directly affect user activity and participation, they do not directly impact other aspects, such as the habits of the moderated community or the way in which the users express themselves. As such, it is safe to assume that all three aforementioned interventions were primarily designed to reduce user activity in problematic subreddits. 
For this specific objective (i.e., reducing activity), all existing studies support the efficacy of Reddit's interventions~\cite{chandrasekharan2017you,saleem2018aftermath,chandrasekharan2020quarantined,horta2021platform} and our new results strongly confirm these findings.

However, a single intervention is capable of affecting multiple dimensions of user behavior~\cite{lazer2018science}, and even interventions designed with a specific objective (e.g., reducing activity) might cause a number of side effects. 
Our results provide evidence that despite achieving the desired objective of reducing the activity of problematic users within and outside of \subr{The\_Donald}, the sequence of interventions also had the \textit{undesired side effects} of increasing the toxicity of such problematic users and of reducing the quality of the news shared and consumed by them (i.e., shared articles became more polarized and less factual). In previous literature, we found mixed support for our findings and many contrasting results. \citeauthor{chandrasekharan2017you} evaluated the effects of banning two hateful subreddits, concluding that former members greatly reduced their hate speech and did not cause an increase in hate speech in the subreddits to which they migrated afterward~\cite{chandrasekharan2017you}. In a subsequent study on two quarantines~\cite{chandrasekharan2020quarantined}, some of the authors from the previous study obtained different results, finding that users in quarantined subreddits did not change their use of toxic terms. They concluded that the quarantine did not serve the goal of reducing offensive posts in the moderated community since it left this dimension essentially unaffected. \citeauthor{horta2021platform} reached yet another conclusion, finding a rise in toxicity following the restriction of two subreddits~\cite{horta2021platform}.

Our results are aligned with findings by \citeauthor{horta2021platform}~\cite{horta2021platform}, since we both measured a significant surge in toxicity following Reddit's interventions. At the same time however, ours also question their interpretation of these findings. \citeauthor{horta2021platform} explained their results in terms of the affordances of social platforms, concluding that the rise in toxicity observed when users migrated to fringe and unregulated platforms could be the «consequence of the removal of platform moderation»~\cite{horta2021platform}. However, in our study we measured a similar rise in toxicity without any reduction in platform moderation, since we studied users who remained on the platform. If anything, our users even faced stronger moderation as they migrated to other subreddits with stricter content policies and rules. An alternative explanation was proposed by \citeauthor{chandrasekharan2020quarantined} in the context of Reddit's quarantines~\cite{chandrasekharan2020quarantined}. They postulated that quarantines could increase the insularity~\cite{allison2020communal} (i.e., polarization and radicalization) of moderated communities and push users toward more extreme positions. Our results seem to confirm this interpretation, since we measured increases in both polarization and toxicity after Reddit quarantined and restricted \subr{The\_Donald}.

The above discussion provides theoretical contributions to the understanding of the effects of moderation interventions. In addition to these, our results also suggest some practical research and policing guidelines. The presence of a multitude of consequences following a moderation intervention mandates care in drawing conclusions on an intervention's efficacy. In particular, future studies should carry out nuanced analyses to assess effects across many behavioral dimensions. In the context of health interventions ---a field from which the study of online moderation interventions inherits many characteristics--- patients that experience side effects of a treatment can provide feedback to their physicians. Here, we lack this valuable feedback and must therefore pay extraordinary attention to the possible unintended consequences of moderation interventions.

\subsection{Implications of Reducing User Activity}
\label{sec:discussion-activity}
To date, all studies that evaluated the effects of Reddit's interventions found evidence for their effectiveness at \textit{reducing the activity} of problematic users. Hence, many studies concluded that such interventions had largely positive effects~\cite{chandrasekharan2017you,saleem2018aftermath,horta2021platform,jhaver2021evaluating}. 
However, their effectiveness depends on the objective that Reddit administrators had when they enforced them, which raises a conundrum. On the one hand, Reddit administrators stated that «one of the primary goals of quarantining is to compel users to rethink their behavior and reduce offensive posts»~\cite{chandrasekharan2020quarantined}, which suggests that their objective was related to \textit{reducing the toxicity} of problematic users. On the other hand however, we noted (\S\ref{sec:discussion-side-effects}) that the mechanics of Reddit's interventions are such that they only directly affect user activity, rather than their toxicity. This reflection surfaces a discrepancy between the motivations stated by Reddit administrators and the moderation interventions that they deployed. Interestingly, the existence of this conceptual discrepancy explains the practical results of our study, where we measured a strong reduction in activity, counterbalanced by an overall increase in toxicity and in the sharing of biased news. More importantly, the discrepancy also suggests that current community-level interventions such as quarantines, restrictions, and bans might be \textit{misdirected} and hence ineffective, or at the very least inefficient, at supporting platform's objectives.

In addition to the above misdirection issue, Reddit's community-level interventions might cause a second problem. Indeed, interventions that reduce activity cause user migrations and, depending on the scope and severity of the intervention, a large number of users might decide to migrate to alternative platforms~\cite{horta2021platform}. Platforms might therefore be disincentivized to apply such interventions, for fear of losing users and revenues~\cite{carlson2020you}. As the only ``solution'' to this issue, current literature appealed to platform's ethical principles, underlining the importance of pursuing the goal of platform moderation even at the cost of reduced revenues~\cite{jhaver2021evaluating}.

The two aforementioned problems highlight some of the limitations of community-level interventions, which appear to be unfit and inappropriate for the current moderation needs of online platforms. For the future, it would thus be advisable to design and deploy nuanced interventions that are more in focus with the objectives of online moderation. To this end, the recent development and experimentation with soft moderation interventions~\cite{zannettou2021won}, as opposed to hard interventions like deplatforming, could be considered as a promising initial step in the right direction. \revision{Then, our findings suggest other possible directions of research, in addition to soft interventions. For instance, the fact that different (groups of) users suffered, in general, different effects to the same intervention ---as in the case of core users of \subr{The\_Donald}--- could motivate the deployment of \textit{diversified} and \textit{personalized} interventions~\cite{cresci2022personalized}. So far, online moderation has always followed a ``one-size-fits-all'' approach, where each intervention was applied in the same way for all users, neglecting individual differences and the advantages of personalization. Instead, for the future we envision the possibility to deploy personalized interventions, each specifically designed for and targeted to a different group of users~\cite{cresci2022personalized}. This novel approach to online moderation will likely boost the effectiveness of moderation interventions, by emphasizing the desired effects of the moderation, while at the same time minimizing the possible undesired side effects.}
Designing nuanced interventions also relates to \citeauthor{kiesler2012regulating}'s theory of graduated sanctions~\cite{kiesler2012regulating}. These could contribute not only at making platforms more accountable for their moderation~\cite{chandrasekharan2020quarantined}, but also at having more targeted interventions, capable of reducing negative side effects such as user migration and loss of revenues. The design and evaluation of targeted and nuanced interventions thus represents fertile ground for future research and experimentation on online moderation.

\subsection{The Current Context for Platform Interventions}
\label{sec:discussion-context}
Since the 2016 Donald Trump presidential win and the unexpected outcome of the UK Brexit referendum, social media platforms have been facing a tremendous pressure to take action against issues such as misinformation and online misbehavior. Recently, the pressure heightened even more as a consequence of other dramatic events such as the George Floyd protests and the emergence of the COVID-19 infodemic. As a result of those events the platforms responded to the growing pressure by hastily enforcing a number of interventions. As notable examples within the scope of our study, Reddit issued several changes to its policies, including those addressing the George Floyd protests murder\footnote{\url{https://redd.it/gxas21}} that eventually led to the ban of \subr{The\_Donald}.\footnote{\url{https://redd.it/hi3oht}} 

Despite appearing as reasonable solutions and serving as public evidence of the platforms' willingness to tackle the issues they contributed to create, many of such interventions were devised and applied light-mindedly. Post-hoc scientific analyses revealed that many interventions proposed in recent years produced mixed effects or no effects at all~\cite{chancellor2016thyghgapp,cheng2014community,chandrasekharan2020quarantined,horta2021platform}, and some even exacerbated the problems they were trying to solve~\cite{bail2018exposure,dias2020emphasizing,pennycook2020implied}. Pennycook and Rand concluded a 2020 op-ed on The New York Times\footnote{\url{https://www.nytimes.com/2020/03/24/opinion/fake-news-social-media.html}} remarking that moderation interventions should «not just rely on common sense or intuition» but that should instead be «empirically grounded». In consideration of our mixed results on the effectiveness of the sequence of interventions on \subr{The\_Donald}, we reiterate this recommendation to balance the timeliness of an intervention with its empirical soundness, motivating present and future research along this important direction.

\subsection{Limitations and Future Work}
\label{sec:discussion-limitations}
\subsubsection{Selection Bias}
On the one hand, our choice of focusing on the sequence of moderation interventions that targeted \subr{The\_Donald} allowed us to thoroughly analyze the former most prominent political community on Reddit. Furthermore, it allowed us to compare and discuss our results with those of several other existing studies that were so far disconnected. For example, we compared our results about the effectiveness of the quarantine and the restriction with the results from~\cite{chandrasekharan2020quarantined} and~\cite{horta2021platform}, respectively. We also compared our results with those of previous works that studied other deplatforming interventions~\cite{chandrasekharan2017you,saleem2018aftermath,jhaver2021evaluating}. Our work thus contributed to reconcile many of the existing studies on the topic.
On the other hand, our analyses are solely focused on the moderation interventions enforced to \subr{The\_Donald} and, as such, might suffer from selection bias and lack of generality. Therefore, more research is needed in order to verify whether our findings hold also for other communities of users, which  could react differently to Reddit's interventions. To this end, our study provides useful guidelines for future research aimed at assessing more precisely the consequences of online moderation strategies.

\subsubsection{Confounders and Exogenous Causes}
An inherent limitation when estimating causal effects with observational data is the susceptibility of the measured quantities to possible exogenous causes (i.e., confounders). Our study, as well as many others~\cite{chandrasekharan2017you,chandrasekharan2020quarantined,horta2021platform,jhaver2021evaluating}, is affected by this limitation. In fact, our analysis did not explicitly consider possible exogenous events that might have contributed to cause some of the effects that we measured. The most noteworthy of such events is the murder of George Floyd, occurred on May 25, 2020, which lays within our post-restriction time window. As such, results about the effects of the restriction might be influenced by this event, as visible in Figure~\ref{fig:rq3_i_toxicity} and as already noted in~\cite{horta2021platform}. To circumvent this limitation, we repeated the BSTS analysis about toxicity changes caused by the restriction, by only considering data up to the killing of George Floyd. The results of this additional analysis confirm our initial findings, in that the restriction caused a marked increase in toxicity. In other words, our conclusions are robust to this exogenous event. Nevertheless, for the future it is advisable to adopt causal inference methodologies that allow to account for, at least, straightforward and known exogenous causes. To this regard, the BSTS method that we adopted to provide our quantitative results admits the possibility to model known exogenous events~\cite{brodersen2015inferring}, and thus represents a favorable methodology for future causal analyses on the effects of online moderation interventions.

\subsubsection{Multidimensionality of Effects}
When elaborating on the possible consequences of the exposure to fake news,~\citeauthor{lazer2018science} concluded that they might have «many potential
pathways of influence, from increasing cynicism and apathy to encouraging extremism», in addition to the obvious consequence of affecting political preferences~\cite{lazer2018science}. In other words, fake news might have \textit{many and diverse effects} spread across multiple dimensions of user behavior and ideology. The same can be said about the interventions that the platforms put in place to contrast fake news and the other ailments that affect online social spaces.
Driven by this consideration, in this work we carried out a nuanced analysis and we moved beyond existing studies that almost solely investigated changes in activity and toxicity (or hate speech)~\cite{chandrasekharan2017you,saleem2018aftermath,chandrasekharan2020quarantined,horta2021platform,jhaver2021evaluating}. We did this by including several important dimensions and metrics that were so far overlooked. Among them is the analysis of the quality of shared news articles, which in turn provides insights into the degree of political polarization and factual reporting of the news consumed by the analyzed communities, and the group community proclivity. Nonetheless, the interventions investigated in our work might have affected many additional dimensions of user behavior and ideology. For example, there might have been changes in user stances toward certain relevant (e.g., political) topics, in their trust of authoritative institutions, and in other emotional, social, and psychological dimensions. Because of this, our results likely provide only a partial view of the full extent of effects caused by the interventions issued on \subr{The\_Donald}. For the future it is thus important to progressively make measurable a larger set of dimensions and to investigate intervention effects also in these dimensions.

\subsubsection{User-level Effects}
Our study considers effects aggregated at the community-level, like most of the works on the same subject~\cite{chandrasekharan2017you,chandrasekharan2020quarantined,horta2021platform,jhaver2021evaluating}. Indeed, aggregated community-level effects are perhaps the most natural and direct way to evaluate community-level interventions (e.g., ban of an entire community). Nonetheless, a community-level post-intervention effect is the combination of many and potentially diverse user-level effects. Hence, the aggregated community-level effect might be weakly representative of the underlying behavior of individuals or smaller user groups. For example, \citeauthor{saleem2018aftermath} measured user-level changes in comment activity and subreddit participation that resulted from the ban of two problematic communities on Reddit~\cite{saleem2018aftermath}, showing very diverse user-level effects.
Based on these considerations, it would be interesting to evaluate user-level effects for a larger set of moderation interventions. In particular, for future work we aim at assessing whether certain community-level effects are the result of homogeneous or heterogeneous user behavior. In the latter case, it would be interesting to investigate which user-level characteristics determine effect diversity, and thus study the possibility of preemptively identifying users likely to significantly deviate from intervention expectations. In either case, understanding effects at the user-level might increase our chances to design and deploy more effective interventions.


%% file: conclusions.tex
\section{Conclusions}
\label{sec:conclusions}
We carried out a multidimensional causal analysis of the sequence of moderation interventions enforced on \subr{The\_Donald}. Our results paint a nuanced picture of the effects of such interventions and support the following take-away messages: (i) the restriction produced stronger effects platform-wise than the quarantine, (ii) core users of \subr{The\_Donald} suffered stronger effects than other users, (iii) both the quarantine and the restriction significantly reduced user activity, however (iv) both also caused an increase in toxicity and (v) caused users to share more polarized and less factual news. We conclude that the sequence of interventions had mixed effects. For the future, it will be important to advance the understanding and the development of moderation interventions, so as to obtain tools capable of achieving the objectives of online moderation with minimal side effects.